\documentclass[journal=jctc,manuscript=article]{achemso}
\usepackage{expl3}
\usepackage{chemgreek}
\setkeys{acs}{maxauthors = 10}
\setkeys{acs}{etalmode=truncate}
\usepackage{graphicx}
\usepackage{hyperref}
\usepackage{amsmath}
\usepackage{amssymb}
\usepackage{braket}
\usepackage{cleveref}
\usepackage{makecell}
\usepackage{diagbox}
\Crefrangeformat{figure}{Fig.~#3#1#4 --~#5#2#6} 
\Crefrangeformat{table}{Tables~#3#1#4 --~#5#2#6} 
\Crefname{figure}{Fig.}{Fig.}
\Crefname{equation}{Eq.}{Eq.}
\creflabelformat{equation}{#2#1#3}
\usepackage{tikz}
\usetikzlibrary{matrix}
\usetikzlibrary{quantikz}
\usepackage{subfig}
\usepackage[version=4]{mhchem}
\usepackage{ragged2e}
\usepackage{multirow}
\usepackage{threeparttable}
\SectionNumbersOn

\title{Efficient measurement schemes for the Monte Carlo projective quantum eigensolver}
\author{Divye Baid}
\author{Maria-Andreea Filip} \email{maf63@cam.ac.uk} 

\affiliation{ Yusuf Hamied Department of Chemistry, University of Cambridge,
Cambridge CB2 1EW, UK }%
\date{\today}
\begin{document}



\begin{abstract}
The Monte Carlo Projective Quantum Eigensolver (MC-PQE) was recently introduced as an alternative to hybrid algorithms like the variational quantum eigensolver (VQE). By using a quantum Monte Carlo-inspired scheme for energy estimation, MC\nobreakdash-PQE was found to decrease the required measurement cost relative to comparable conventional approaches. However, in the context of VQE, numerous techniques have been developed to reduce the measurement overhead by joint measurement of multiple observables. In this work, we extend these approaches to the asymmetric expectation values required in MC-PQE and assess the performance of different techniques for various molecular systems of up to 12 qubits. We find full commuting Pauli term grouping combined with tailored measurement allocation techniques leads to a 5-10$\times$ reduction in standard error for the same total number of quantum measurements. Conventional Hamiltonian grouped measurement outperforms tractable classical shadows tomography based techniques for the considered systems.
\end{abstract}
\maketitle

\section{Introduction}
Quantum computers were proposed as a hypothetical simulation paradigm for quantum physical systems by Feynman in the 1980s.\cite{Feynman1982} In the past decade, quantum computing technology has advanced significantly, with multiple different physical qubit technologies developing in parallel\cite{Bartee2025SpinQubitCMOS,King2025BeyondClassical,PsiQuantum2025Manufacturable,Steinacker2025IndustryCompatible,Bluvstein2026FaultTolerant} and promising error correction experiments.\cite{Tan2025,GoogleQuantumAI2025BelowThreshold,Caune2026RealTimeQEC,Drouet2026} Nevertheless, noise remains a significant challenge in quantum computation, as the field slowly transitions from noisy intermediate scale quantum devices\cite{Preskill2018} to error-corrected fault-tolerant machines.

In order to make use of current devices, many hybrid quantum algorithms have been devised, which off-load only a few crucial operations to a quantum computer, while still carrying out the majority of the computation classically. The first of these with applications to quantum chemistry and many-body physics systems was the Variational Quantum Eigensolver (VQE),\cite{Peruzzo2014} which uses classical optimisation of a cost function calculated by a quantum processor to find approximate ground state wavefunctions. VQE has since been used for a variety of chemistry applications.\cite{OMalley2016,Kandala2017,Shea2017,Hempel2018,Grimsley2019b,Tilly2020,Sugisaki2022,Clary2023,Dalton2024,Kim2024,Alfonso2025,Weaving2025} However, optimisation of quantum wavefunctions has proved to be challenging\cite{McClean2018,Cerezo2021,Wang2021,Choy2025} and VQE often incurs an intractable measurement cost, due to the many repetitions needed to measure quantum expectation values to sufficient accuracy.

Various other hybrid schemes have been developed to avoid this measurement overhead. A currently popular effort is quantum selected configuration interaction (QSCI)\cite{Kanno2026} or sample-based quantum diagonalisation (SQD),\cite{Shajan2025} which only uses a quantum device to sample important contributions to the wavefunction and then performs a classical diagonalisation in the space these configurations define. While the original SQD algorithm was Ansatz-based, an alternative employing a time-evolution Krylov subspace has also been developed.\cite{Piccinelli2026} This type of approach is measurement-efficient, but it encounters a classical bottleneck as the selected configuration interaction (SCI) space often grows exponentially with problem size and SQD sampling is often redundant.\cite{Reinholdt2025}

Another direction has been to retain an optimised wavefunction Ansatz, obtained however by projective rather than variational techniques. The projective quantum eigensolver (PQE)\cite{Stair2021,Misiewicz2024} employed a linear ground-state projector, similar to that used in diffusion Monte Carlo (DMC)\cite{Reynolds1990} or full configuration interaction quantum Monte Carlo (FCIQMC),\cite{Booth2009} to obtain optimal Ansatz parameters. While this is more reliable than conventional VQE, the measurement cost is still prohibitive. Recently, one of us proposed a quantum Monte Carlo-inspired stochastic version of PQE, which showed significant measurement overhead reductions, while maintaining accuracy.\cite{Filip2024}

In the context of VQE, a variety of measurement techniques have been devised to decrease the overhead of expectation value evaluation.\cite{Patel2025} In this work, we adapt these schemes to the Monte Carlo PQE (MC-PQE) algorithm, which requires asymmetric expectation values. 

\Cref{sec:background} gives an overview of the measurement techniques developed for expectation value evaluation and of the MC-PQE algorithm. In \Cref{sec:schemes} we analyse the propagation of individual observable variances in the MC-PQE algorithm and present adapted measurement schemes. We evaluate the performance of these schemes for different molecular systems in \Cref{sec:results} and provide conclusions in \Cref{sec:conclusions}.

\section{Background theory}\label{sec:background}
\subsection{Measurement techniques}
There are two types of measurement techniques which have been used to reduce the sampling overhead in hybrid quantum algorithms. The first is operator term grouping,\cite{Patel2025} which aims to measure contributions from multiple Pauli operators simultaneously, and the other is classical shadow tomography,\cite{Huang2020,Huang2021,Hadfield2022} where measurements are carried out in different randomised bases to construct an approximation to the quantum density matrix.
\subsubsection{Hamiltonian Grouping}
Consider a Hamiltonian $\hat H = \sum_k h_j \hat P_k$, where $\hat P_k$ are strings of Pauli operators. In order to obtain the expectation value of $\hat H$, one must obtain expectation values of all $\hat P_k$. If $[\hat P_k,\hat P_j] = 0$, then the expectation values of both can be measured simultaneously. Most Hamiltonian grouping techniques are based around forming groups of mutually commuting operators, which may overlap\cite{Choi2022,Wu2023,Choi2023} or not.\cite{Verteletskyi2020,Gokhale2020,Yen2020,Huggins2021,Yen2021,Crawford2021,Gena2022} There have also been some attempts to group anti-commuting Pauli terms together.\cite{Izmaylov2020} In this work, we will focus on two approaches to obtain non-overlapping commuting groups: qubit-wise commutativity and full commutativity implemented by sorted insertion.\cite{Crawford2021}

{\textit{Qubit-wise Commutativity.}}
Qubit-wise commutativity (QWC) is the simplest Hamiltonian grouping technique, relying on the fact that Pauli matrices commute with themselves $[X,X]=[Y,Y]=[Z,Z]$ and with the identity $[X,I]= [Y,I] = [Z,I] = 0$, and enforcing commutativity between the Pauli matrices acting on each qubit. One approach to obtain QWC groups, employed in the initial MC-PQE implementation, is as follows:

\begin{enumerate}
    \item {Let $T$ be the tuple initially containing all $k$ Pauli strings with coefficients $c_i P_i$.}
    \item {For each Pauli string in $T$, its shape is defined as the phaseless string obtained by replacing each instance of a particular Pauli operator (either $X$, $Y$, or $Z$) with $I$. Each collection is the set of all Pauli strings with the same shape.}
\end{enumerate}

The exact choice of Pauli operator in step 2 may lead to different QWC partitionings. For example, $\{XX, IX, ZX \}$ will be partitioned into $\{ XX, IX \}$, $\{ZX \}$ using $X$-based commutativity, or $\{XX \}, \{IX, ZX \}$ using $Z$-based commutativity. In this work, we choose $Z$-based commutativity as convention. Thus, applied to typical Hamiltonians, QWC yields one large collection of diagonal terms (strings of only $I$ or $Z$), and many small collections mostly with one term each. For example, the Hamiltonian of $\text{H}_4$ contains 185 Pauli strings, and QWC yields one diagonal-only collection of 37 strings, eight collections of 7 strings and 56 collections of one string each. A simple analysis shows that QWC on a Hamiltonian containing $\mathcal{O}(n^4)$ $n$-qubit Pauli strings yields $\mathcal{O}(n^3)$ collections.\cite{Verteletskyi2020,Yen2023} For the full set of $4^n$ $n$-qubit phaseless Pauli strings, $\mathrm{P}_n / \mathrm{U}(1)$, applying QWC yields exactly $3^n$ collections.\cite{Kurita2023}

{\textit{Sorted Insertion.}} QWC can be improved on by grouping Pauli terms into fully commuting groups. Unlike QWC, full commuting group methods consider only whether the overall Pauli strings commute and will therefore group strings with even numbers of anti-commuting single-qubit Paulis together. In general, this is an NP-hard problem,\cite{Yen2020} although efficient heuristic algorithms exist.\cite{Verteletskyi2020} One example is sorted insertion (SI),\cite{Crawford2021} which preferentially collects Pauli strings with large coefficients into large collections. For a Hamiltonian containing $k$ distinct Pauli strings, its procedure is as follows:

\begin{enumerate}
    \item {Let $T$ be the tuple initially containing all $k$ Pauli strings $c_i P_i$, sorted in descending order of $|c_i|$.}
    \item {For the first Pauli string $c' P'$ in $T$, check if it commutes with every member of an existing collection (in order of creation of the collections). Remove $c' P'$ from $T$ and place it in the first such collection; if no such collection exists place it in a new collection.}
    \item {Repeat step 2 until $T$ is empty.}
\end{enumerate}

It has been shown that the optimal (smallest) number of collections required to cover $4^n$ Pauli strings is $2^n + 1$.\cite{Bandyopadhyay2002,Kurita2023} This immediately demonstrates an improvement over QWC. Generating collections of commuting groups through heuristic algorithms is not guaranteed to generate the optimal collections, but in practice it is found to often outperform QWC grouping.\cite{Yen2023} 

\subsubsection{Classical Shadow Tomography}
So-called ``classical shadows" have been proposed as an efficient way of estimating multiple observables simultaneously.\cite{Huang2020} The approach works as follows, given a quantum state $\ket{\Psi}$, with density $\rho = \ket{\Psi}\bra{\Psi}$:
\begin{enumerate}
    \item Select a unitary operator $U_k$ with uniform probability from a fixed ensemble $\mathcal{U}$ and apply it to $\ket{\Psi}$.
    \item Measure in the computational basis to obtain basis state $b_k$.
    \item Store $U_k^\dagger\ket{b_k}\bra{b_k}U_k$.
\end{enumerate}
The expectation value over these stored quantities represents a linear map of the initial density
\begin{equation}
    \mathcal{M}(\rho) = \mathbb{E}_k[U_k^\dagger\ket{b_k}\bra{b_k}U_k]
\end{equation}

If this map can be inverted, it is possible to obtain a representation of the original density as
\begin{equation}
    \rho = \mathbb{E}_k[\hat \rho_k] = \mathbb{E}_k[\mathcal{M}^{-1}(U_k^\dagger\ket{b_k}\bra{b_k}U_k)],
\end{equation}
as well as operator expectation values as
\begin{equation}
    \braket{\hat O} = \mathbb{E}_k[\text{Tr}[\hat O \hat \rho_k]].
\end{equation}
Commonly used ensembles include the $n$-qubit Clifford group and the $n$-qubit Pauli group, which give invertible maps and
\begin{equation}
    \hat \rho^\text{Clifford} = (2^n + 1) U^\dagger\ket{b}\bra{b}U- I,
\end{equation}
while
\begin{equation}
    \hat \rho^\text{Pauli} = \otimes_{i=1}^n \left(U_i^\dagger\ket{b_i}\bra{b_i}U_i - I\right),
\end{equation}
where in this case $U = \otimes_{i=1}^n U_i$, with $U_i$ single qubit Pauli operators. It has been shown that the number of measurements needed to obtain a fixed additive error in all of the operators considered scales logarithmically with the number of operators.\cite{Huang2020} So, in the case of the $\mathcal{O}(n^4)$ terms in the molecular Hamiltonian, this approach would only require $\mathcal{O}(\log(n))$ shots.

\subsection{Monte Carlo Projective Quantum Eigensolver}

Given a wavefunction $\Psi(0)$ with non-zero overlap with the true ground state $\Psi_0$ of a Hamiltonian $\hat H$, with energy $E_0$,
\begin{equation}
    \ket{\Psi_0} \propto \lim_{\beta \rightarrow\infty} e^{-\beta (\hat H-E_0)}\ket{\Psi(0)},
\end{equation}
where $\beta$ is imaginary time. For small imaginary time-steps $\Delta \beta = \beta/n$, this may be rewritten as
\begin{equation}
    \ket{\Psi_0} = \lim_{n \rightarrow \infty} {[1 - \Delta\beta (\hat H - E_0)]}^n \ket{\Psi(0)}.
    \label{eq:proj}
\end{equation}

Conventional projective quantum Monte Carlo algorithms\cite{Booth2009,Thom2010,Filip2020} and MC-PQE use an iterative form of this equation,
\begin{equation}
        \ket{\Psi(\beta + \Delta \beta} = {[1 - \Delta\beta (\hat H - E_0)]} \ket{\Psi(\beta)},
\end{equation}
which can be projected onto the various Slater determinants in the Hilbert space to give
\begin{equation}
    \braket{\Phi_i|\Psi(\beta + \Delta \beta)} = \braket{\Phi_i|{[1 - \Delta\beta (\hat H - E_0)]}|\Psi(\beta)}.
    \label{eq:pmc}
\end{equation}
MC-PQE requires a parametrisation of $\ket{\Psi(\beta)} = U(\boldsymbol{\theta}(\beta)) \ket{\Phi_0}$, where $\Phi_0$ is an initial reference state, normally the Hartree--Fock determinant, and $U(\boldsymbol{\theta}(\beta))$ is a unitary wavefunction Ansatz, such that for every parameter $\theta_i$ in the set $\boldsymbol{\theta}$ there exists a distinct Slater determinant for which
\begin{equation}
    \braket{\Phi_i|\Psi(\beta)} = \theta_i + \mathcal{O}(\theta^2).
\end{equation}
For the analysis in this paper, we will employ the unitary coupled cluster singles and doubles (UCCSD) Ansatz.\cite{Kutzelnigg1982,Kutzelnigg1983,Kutzelnigg1984,Bartlett1989} However, the MC-PQE is compatible with other methodologies, such as Givens-rotation based wavefunctions.\cite{Arrazola2022,Rajan2025,Welling2026} In UCCSD, the wavefunction is given by
\begin{equation}
    \ket{\Psi_\text{UCCSD}} = e^{\hat \tau} \ket{\Phi_0},
\end{equation}
where $\hat \tau = \sum_i \theta_i(\hat a_i - \hat a_i^\dagger)$,
where $\hat a_i$ are particle-hole single and double excitation operators.

In MC-PQE, the wavefunction is represented by a population of ``walkers" on each parameter, as in conventional QMC algorithms.\cite{Booth2009,Thom2010,Filip2020} This allows a sparse representation of the wavefunction, where only parameters with significant coefficients are stored a meaningful percentage of the time. The evolution of this population follows a dynamics similar to conventional QMC, the details of which may be found elsewhere.\cite{Filip2024} For our purposes, it is important to note that a quantum computer is used to measure the residual
\begin{equation}
    r_i(\beta) = \braket{\Phi_i|{\Delta\beta (\hat H - S)}|\Psi(\beta)} = \Delta\beta(H_{i} - S c_{i}),
    \label{eq:res}
\end{equation}
where $S$ is a population control parameter which converges to the ground-state energy when the calculation has reached steady-state, $H_{i} = \braket{\Phi_i|H|\Psi}$ and $c_{i} = \braket{\Phi_i|\Psi}$. The computation of both $H_i$ and $c_i$ may be achieved through a modified Hadamard test.\cite{Filip2024} The populations of each parameter $N_i$ are then updated as
\begin{equation}
    N_i(\beta + \Delta\beta) = N_i(\beta) - N_0(\beta) r_i,
\end{equation}
where $N_0$ is the population on the reference determinant. The energy is computed by the projected energy estimator,
\begin{equation}
    E_\mathrm{proj} (\beta) = \frac{\braket{\Phi_0|\hat H| \Psi(\beta)}}{\braket{\Phi_0|\Psi(\beta)}} = \frac{H_0}{c_0}.
    \label{eq:e_proj}
\end{equation}

The shift is usually initialised with $S = E_\mathrm{HF}$, the energy of the reference determinant. In practice, QMC algorithms tend to use a shifted Hamiltonian $\hat K = \hat H - E_\text{HF}$, so the initial $S=0$. This generates more stable dynamics due to the smaller magnitude of most diagonal Hamiltonian terms. In this regime, the total population $N_w = \sum_i |N_i|$ will grow exponentially. Once a threshold population has been reached, the shift is set to the current $E_\mathrm{proj}$ and then updated as
\begin{equation}
        S(\beta) = S(\beta - A\delta\beta) - \frac{\zeta}{A\delta\beta}\mathrm{ln}\frac{N_w(\beta)}{N_w(\beta - A\delta\beta)}.
        \label{eq:shift}
    \end{equation}
This maintains the population approximately constant and allows $S$ to function as an independent estimator for the ground state energy. Once the steady-state regime is reached, the final estimate of the ground state energy is obtained by a reblocking analysis\cite{Flyvbjerg1989} of the shift and projected energy estimators. This approach is what leads to the significant decrease in measurement cost compared to VQE or PQE, which require highly accurate individual values for their energy estimators. However, further gains can be obtained by stochastically sampling the wavefunction and the Hamiltonian at each step, reducing circuit complexity and observables to be measured respectively.

In the following section, we derive the dependence of the final variance of the shift and projected energy estimators on the variance of the individual $H_i$ and $c_i$ measurements at each step. We will then devise measurement schemes aiming to reduce the variance of these estimators, using both simple Hamiltonian grouping techniques and classical shadows.
\section{Measurement Schemes for MC-PQE}\label{sec:schemes}

\subsection{Variance propagation in MC-PQE}

In VQE, the final energy estimator is a weighted sum of the quantities measured on the quantum device, so the variance of these quantities propagates directly to the estimator. In MC-PQE, the relation between the estimators and the measured quantities is more complex. 

At each iteration, we obtain quantum estimates for $H_i$ and $c_i$, with associated variance $\sigma_{H_i}^2$ and $\sigma_{c_i}^2$. We now consider the relationship between these and the final shift $\bar S$ and projected energy $\bar E_\text{proj}$ estimators.

For the purposes of the discussion here, we assume the MC-PQE estimators are simple means of the shift and projected energy values obtained in the steady-state phase of the calculation. By ignoring the autocorrelation time and blocking procedure, we obtain lower bounds on the true variances of these estimators. We also assume $H_i$ and $c_i$ are measured from independent circuits, so $\text{Cov}(H_i, c_j) = 0,\ \forall i,j$ and that, in the steady-state regime, $\sigma_{c_i}^2$ and $\sigma_{H_i}^2$ are the same at every time-step. 

\textit{Projected energy}. The projected energy is estimated according to \Cref{eq:e_proj} at each step. However, in practice the final estimator is computed as
\begin{equation}
    \bar E_\text{proj} = \frac{\bar H_0}{\bar c_0},
\end{equation}
where $\bar H_0$ and $\bar c_0$ are the averages of the respective quantities over the $N$ steady-state iterations. Therefore, the variance is given by
\begin{equation}
    \frac{\sigma^2_{\bar E_\text{proj}}}{\bar E_\text{proj}^2} = \frac{\sigma^2_{\bar H_0}}{\bar H_0^2} + \frac{\sigma^2_{\bar c_0}}{\bar c_0^2} = \frac{1}{N}\left(\frac{\sigma^2_{H_0}}{\bar H_0^2} + \frac{\sigma^2_{ c_0}}{\bar c_0^2}\right),
\end{equation}
where $N$ is the number of steps the expectation value is accumulated over. As expected, the noise in the projected energy estimator is dominated by the noise in the Hartree--Fock overlap and Hamiltonian term.

\textit{Shift.} The variance of the shift estimator is more complex, due to the update equation (\Cref{eq:shift}). We consider the relative quantity
\begin{equation}
    S_r(\beta) = S(\beta) - E_0,
\end{equation}
which is the term that induces changes to the total populations in the steady-state regime. Then, to first order
\begin{equation}
    N_w(\beta + A\delta\beta) = N_w(\beta) (1 +  S_r(\beta)A\delta\beta) + \epsilon_N(\beta),
\end{equation}
where $\epsilon_N(\beta)$ is the noise term due to the variance in the measured quantum observables.Therefore,
\begin{equation}
    \ln\left(\frac{N_w(\beta+A\delta\beta)}{N_w(\beta)}\right) = \ln(1+S_r(\beta)A\delta\beta + \epsilon(\beta)) \approx S_r(\beta)A\delta\beta + \epsilon(\beta).
\end{equation}
where $\epsilon(\beta) = \epsilon_N(\beta)/N_w(\beta)$ is the noise in the population ratio.
Therefore, the shift becomes approximately
\begin{equation}
    S(\beta +A\delta\beta) = S(\beta)(1-\zeta) + \frac{\zeta}{A\delta\beta}\epsilon.
\end{equation}
If $\epsilon$ is independent of $\beta$, this is an auto-regressive model of order 1 (AR(1)),\cite{ShumwayStoffer2017} for which there exist standard variance results. We find that while this assumption does not strictly hold, it provides useful insight into the relationship between the variance of the shift and that of the measured quantum variables. We note this treatment is similar to that employed in Ref. \citenum{Yang2020,Brand2022} to understand population control and the bias it introduces in FCIQMC, although they take a continuous differential equation approach.

For a series of random variables $X_{n+1} = \varphi X_n + \epsilon'$, the variance of the variable at each step is given by
\begin{equation}
    \sigma_{X}^2 = \frac{\sigma_{\epsilon'}^2}{1 - \varphi^2}
\end{equation}
and the variance of the mean is given by
\begin{equation}
    \sigma_{\bar X}^2 = \frac{\sigma_{X}^2}{N}\frac{1+\varphi}{1-\varphi}.
\end{equation}
In the case of the shift, this becomes
\begin{equation}
    \sigma_{\bar S}^2 = \frac{\sigma_S^2}{N}\frac{2-\zeta}{\zeta} = \frac{\sigma_\epsilon'^2}{N \zeta^2},
\end{equation}
which is well-defined for the normally employed values of $\zeta \in [0.01,0.1]$. Since $\epsilon' = \frac{\zeta}{A\delta\beta}\epsilon$, $\sigma_\epsilon'^2 = \frac{\zeta^2}{A^2\delta\beta^2}\sigma_\epsilon^2$, so overall
\begin{equation}
    \sigma_{\bar S}^2 = \frac{\sigma_\epsilon^2}{N A^2(\delta\beta)^2}.
\end{equation}
The noise in the population estimator can then be related to the variance of the $H_i$ and $c_i$ quantities as follows.

The total walker population at time $\beta +\delta\beta$ is given by
\begin{equation}
    N_w(\beta +\delta\beta) = N_0\sum_i|\theta_i(\beta +\delta\beta))| = N_0\sum_is_i(\beta +\delta\beta)\theta_i(\beta +\delta\beta)),
\end{equation}
where $s_i$ is the sign of the amplitude $\theta_i$. In what follows, we will assume this to be independent of $\beta$, which is reasonable in the steady-state regime. Therefore
\begin{equation}
\begin{split}
    N_w(\beta +\delta\beta) &=N_0(\beta)\sum_is_i\{\theta_i(\beta) - \delta\beta[H_i(\beta) +\epsilon_{H_i} - S(\beta) (c_i(\beta)+\epsilon_{c_i})]\} \\
    &= N_w(\beta) - N_0(\beta)\delta\beta\sum_i s_i[H_i(\beta) +\epsilon_{H_i} - S(\beta) (c_i(\beta)+\epsilon_{c_i})]\}.
\end{split} 
\end{equation}
The noise in $N_w$ is then
\begin{equation}
    \epsilon_N(\beta) = N_0(\beta)\delta\beta\sum_i s_i(\epsilon_{H_i} - S(\beta)\epsilon_{c_i})
\end{equation}
and
\begin{equation}
    \epsilon(\beta) = \frac{\epsilon_N(\beta)}{N_w(\beta)} = \frac{\delta\beta\sum_i s_i(\epsilon_{H_i} - S(\beta)\epsilon_{c_i})}{\sum_i s_i \theta_i(\beta)}.
\end{equation}
Further assuming that the cluster amplitudes are approximately constant in the steady state gives
\begin{equation}
    \sigma_\epsilon^2(\beta) = \frac{\delta\beta^2\sum_i(\sigma_{{H_i}}^2 + S(\beta)^2\sigma_{c_i}^2)}{\left(\sum_is_i \theta_i\right)^2}.
\end{equation}
This violates the constant $\sigma_\epsilon^2$ assumption of the AR(1) model, as the variance depends on $S(\beta)$. Once again, around steady-state we can assume $S(\beta) \approx \bar S$, so the model is a valid approximation. Finally, we find that
\begin{equation}
    \sigma_{\bar S}^2 = \frac{\sum_i(\sigma_{H_i}^2 + \bar{S}^2\sigma_{c_i}^2)}{N A^2\left(\sum_i s_i \theta_i\right)^2}.
    \label{eq:shift_noise}
\end{equation}
Therefore, while the variance of the projected energy estimator only depends on the variance of the Hartree--Fock Hamiltonian term, the shift depends on a sum of all measurements. This justifies the observed trend that shift variances tend to be larger than those of the projected energy for the same system. From this analysis, it is clear that reducing the variances of both the overlap and Hamiltonian terms is crucial to reduce the noise in the MC-PQE estimators and therefore the measurement overhead.
\subsection{Hamiltonian grouping}

The original MC-PQE results were obtained using QWC Hamiltonian grouping, with an equal measurement split between all groups. In this work, we propose improvements to both of these aspects. We employ sorted insertion to reduce the number of Hamiltonian groups, with results for typical molecular systems presented in \Cref{tab:grouping}. We note that for a Hamiltonian of the form $\hat H = \sum_i h_k \hat P_k$, the Paulis that need to be measured in MC-PQE are $\hat Z_\text{anc}\otimes \hat P_k$, so the groups are the same as those resulting from the original Hamiltonian, as the ancilla operator always commutes with itself.

\begin{table}[]
    \centering
    \begin{tabular}{|c|c|c|c|c|}
    \hline
      \textbf{System }  & \textbf{QWC Groups} & \textbf{SI Groups }& \textbf{QWC Observables} & \textbf{SI Observables}\\
      \hline
        H$_4$ & 101 & 15 & 1530 & 165 \\
        LiH & 125& $19-21$&1386 &$220-242$\\
        BeH$_2$ &313 & $26-33$& 9734&$837-1054$\\
        H$_2$O & 313& $29-33$& 9734&$930-1054$\\
        HF & 125& $20-21$&1386 &$231-242$\\
        \hline
    \end{tabular}
    \caption{Number of Pauli groups in the molecular Hamiltonian and total number of MC-PQE observables measured for the systems considered in this work.}
    \label{tab:grouping}
\end{table}

Once SI groups are determined, we can keep the uniform measuring scheme or introduce a measurement allocation procedure.\cite{Yen2023} It has been shown that for VQE, for a fixed total number of shots, the optimal measurement allocation gives
\begin{equation}
    m_G = M \frac{\sqrt{\text{Var}(H_G)}}{\sum_G\sqrt{\text{Var}(H_G)}},
\end{equation}
where $H_G = \sum_{j \in G} h_j\braket{\Psi|P_j|\Psi}$. 

Unlike VQE, the measured quantities in MC-PQE are not simply added together into an expectation value, which makes allocation non-trivial. Is is easy to see that for each $H_i = \sum_{G}\sum_{j\in {G}} h_j\braket{\Phi_i|\hat P_j|\Psi}$, measurements should be allocated according to the variance of each group $G$. However, the question then becomes the allocation of measurements to different $H_i$ and $c_i$. We propose the following options:
\begin{enumerate}
    \item \textbf{Pure variance (PV):} allocate the total number of measurements $M$ according to the individual variances of the different quantities, as 
    \begin{align}
        m_{h_{iG}} = M\frac{\sqrt{\text{Var}(h_{iG})}}{\sum_i\left(\sum_G\sqrt{\text{Var}(h_{iG})} +\sqrt{\text{Var}(c_i)}\right)}\\
        m_{c_{i}} = M\frac{\sqrt{\text{Var}(c_{i})}}{\sum_i\left(\sum_G\sqrt{\text{Var}(h_{iG})} +\sqrt{\text{Var}(c_i)}\right)},
    \end{align}
    where $h_{iG} = \sum_{j\in {G}} h_j\braket{\Phi_i|\hat P_j|\Psi}$.
    \item \textbf{Shift-weighted (SW):} allocate the total number of measurements $M$ to minimise the variance of the shift (\Cref{eq:shift_noise}), as
        \begin{align}
        m_{h_{iG}} = M\frac{\sqrt{\text{Var}(h_{iG})}}{\sum_i\left(\sum_G\sqrt{\text{Var}(h_{iG})} +S\sqrt{\text{Var}(c_i)}\right)}\\
        m_{c_{i}} = M\frac{S\sqrt{\text{Var}(c_{i})}}{\sum_i\left(\sum_G\sqrt{\text{Var}(h_{iG})} +S\sqrt{\text{Var}(c_i)}\right)}
    \end{align}
    \item \textbf{Even-variance (EV):} allocate measurements evenly between different $H_i$ and $c_i$ and then by variance within each $H_i$.
    \item \textbf{Even-determinant (ED):} allocate measurements evenly between determinants and then use either PV or SW within each determinant.
\end{enumerate}
The choice among these options is not obvious, as (2) requires knowledge of the shift and none are necessarily optimal for both shift and projected energy. We will explore all three in \Cref{sec:results}.

We note that in practice, terms like $\braket{\Phi_i|\hat P_j|\Psi} = \braket{\Psi_a|\hat Z_a \otimes \hat P_j|\Psi_a}$, where
\begin{equation}
    \ket{\Psi_a} = \frac{1}{2}((\ket{0}+\ket{1})\ket{\Phi_i} + (\ket{0}-\ket{1})\ket{\Psi})
\end{equation}
and $\hat Z_a$ is the $Z$ operator on the ancilla in the Hadamard test. Therefore, the variance is well defined and can be computed from the same $\ket{\Psi_a}$.

An additional problem is posed by the wavefunction used to compute the variance of the different terms. Yen \textit{et al.}\cite{Yen2023} use full configuration interaction (FCI) as a benchmark and propose configuration interaction singles and doubles (CISD) as a heuristic or an iterative measurement allocation scheme (IMA) to make use of the true VQE wavefunction. As IMA introduces a significant overhead, we consider simpler options here:
\begin{enumerate}
    \item use the initial Hartree--Fock wavefunction for measurement allocation;
    \item use the initial Hartree--Fock wavefunction for measurement allocation, then recompute with the current wavefunction when the shift starts to vary;
    \item use the FCI wavefunction for benchmarking.
\end{enumerate}

\subsection{Shadow tomography}

Recently, Lenihan \textit{et al.}\cite{Lenihan2025} employed partial shadow tomography to compute the projected energy, calculated as
\begin{equation}
    E_\mathrm{proj} = E_\text{HF}+\frac{1}{c_0} \sum_{\substack{i,j \in \text{occ}\\a,b\in\text{virt}}}c_{ij}^{ab} \braket{\Phi_0|\hat H|\Phi_{ij}^{ab}}.
\end{equation}
The coefficients $c_0$ and $c_{ij}^{ab}$ can be extracted by state tomography. In order to do so, they define a density operator $\tilde \rho = \ket{\tau}\bra{\tau}$, where
\begin{equation}
    \ket{\tau} = \frac{1}{\sqrt{2}}(\ket{0} + \ket{\Psi})
\end{equation}
and operators
\begin{equation}
    \hat O_{ij}^{ab} = \ket{0}\bra{\Phi_{ij}^{ab}} + \ket{\Phi_{ij}^{ab}}\bra{0},
\end{equation}
which makes
\begin{equation}
    c_{ij}^{ab} = \text{Tr}[\hat O_{ij}^{ab}\tilde\rho].
\end{equation}
The quantities can therefore be computed as\cite{Lenihan2025}
\begin{equation}
    c_{ij}^{ab} = 2 (2^N + 1)\mathbb{E}_k[\braket{\Phi_{ij}^{ab}|\hat U_k|b_k}\braket{b_k|\hat U_k|0}],
    \label{eq:c_exp}
\end{equation}
where $\hat U_k$ are $n$-qubit Cliffords and $b_k$ are the sampled bit-strings as part of the classical shadows procedure. This procedure can be trivially expanded to contributions beyond double excitations. In order to compute the residuals in \Cref{eq:res} for the UCCSD Ansatz, one would need to obtain amplitudes up to quadruple excitations. The number of shots required to get a fixed additive precision scales logarithmically with system size.\cite{Lenihan2025} Additionally, unlike the Hadamard test employed above, this approach requires no ancilla qubits or controlled excitation operations, significantly decreasing the cost of each circuit evaluation. We investigate how this compares to conventional Hamiltonian measurement at practical shot counts.

\section{Results}\label{sec:results} 
Calculations are carried out using an in-house MC-PQE code, developed using the pytket package.\cite{Sivarajah2021} All quantum measurements are simulated on a classical computer. In all cases, MC-PQE calculations are run to a stable population threshold and 2000 steady-state iterations are used to estimate the shift and projected energy, rather than varying this window to converge estimators to particular thresholds. We employ this routine for consistency, in order to be able to study the effect of different simulation algorithms across different systems. We note that it is not necessarily sufficient to achieve sub-milliHartree standard errors in all cases and results should only be compared relative to one another.

\subsection{Hamiltonian Grouping}
We consider the grouping options proposed above for different molecular systems, including H$_4$ and first row hydrides. We note that the SW method would allocate no measurements to the overlaps before the shift starts varying, so cannot be employed with the pure Hartree\nobreakdash--Fock variance. We consider a combination where the variance is computed with the PV method, then recomputed once the shift begins varying, without also updating the underlying wavefunction used to compute the variances. 

\begin{table}[]
    \centering
    \begin{threeparttable}
        \caption{Standard error in mHartree of observables computed from MC-PQE runs with evenly distributed measurements for H$_4$/STO-3G, with $r_\text{HH}=1.5$\AA.}
        \label{tab:h4_1}
         \begin{tabular}{|c|c|c|c|c|c|}
         \hline
         \textbf{Observable} & \makecell{\textbf{QWC}\\$N_s = 10^5$}  & \makecell{\textbf{QWC}\\$N_s = 1.53 \times 10^5$$^a$} &\makecell{\textbf{QWC}\\$N_s = 10^6$} & \makecell{\textbf{SI}\\$N_s = 10^5$}&\makecell{\textbf{SI}\\$N_s = 10^6$} \\
         \hline
       {${S}$}  & 1.49 & 1.10& 0.60 &0.56&0.22\\
        \hline 
        {${E_\text{proj}}$} & 3.75 & 2.70 & 1.26 &1.57&0.50\\
        \hline
    \end{tabular}
    
    \begin{tablenotes}
        \footnotesize
        \item[a] $N_s = 1.53\times 10^5$ shots for QWC allocate 100 shots/observable, which is the minimum allowed in an SI calculation.
    \end{tablenotes}
    
    \end{threeparttable}
   
\end{table}

Results for H$_4$/STO-3G at $r_\text{HH} = 1.5$\AA\ are tabulated in \Cref{tab:h4_1,tab:h4_2}. \Cref{tab:h4_1} compares the standard deviations of the shift and projected energy estimators obtained with the even measurement split, using QWC and SI. Results are obtained as $\sigma = \sqrt{\frac{\sum_{k=1}^K \sigma_{\bar O}^2(k)}{K}}$ over $\mbox{K=5}$ independent trials. As expected, SI produces lower standard errors for the same shot count. Further dependence on the shot distribution is shown in \Cref{tab:h4_2}. We find notable improvement from QWC to SI results, with an approximate $10\times$ reduction in the number of shots needed to maintain the same standard error in SI. Looking at the different shot distribution methods, we find that some outperform even splitting, although for this system the effect is relatively small and the best method differs between the shift and projected energy. In all cases, using a correlated wavefunction as the variance source significantly outperforms the HF wavefunction, with the HF + QMC method very close to the FCI. This is to be expected as, in H$_4$, the UCCSD wavefunction has very high overlap to the FCI ground state.

\begin{figure}
     \centering
     \includegraphics[width=\linewidth]{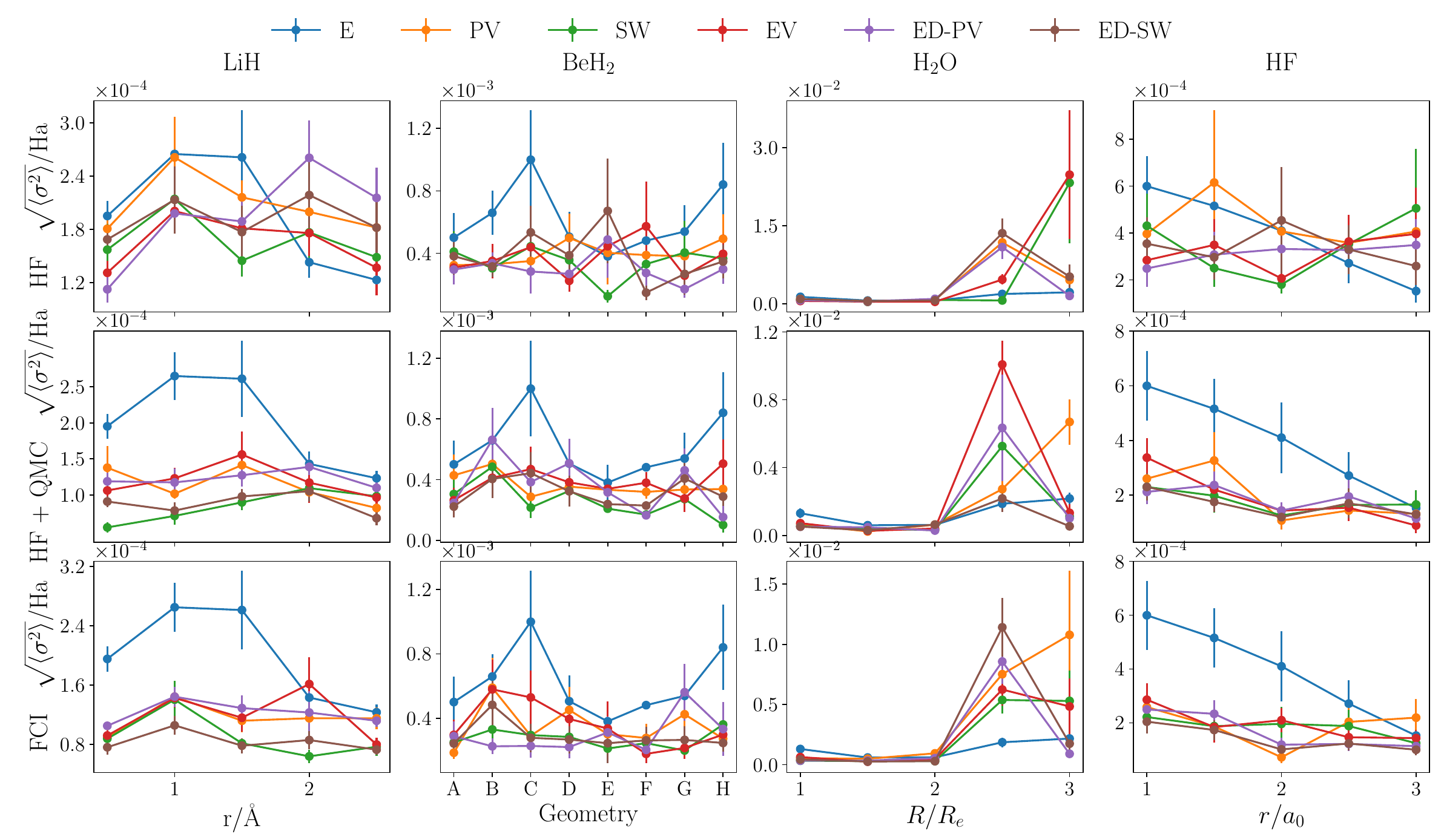}
     \caption{Average standard error in the shift obtained by MC-PQE, using different measurement allocation techniques and estimated variance sources. The number of samples is $N_s = 10^6$ for LiH, BeH$_2$ and H$_2$O and $N_s = 10^7$ for HF. Error bars are obtained from 5 independent simulations. The even allocation (E) is independent of variance source.}
     \label{fig:vars_shift}
 \end{figure}

\begin{figure}
    \centering
    \includegraphics[width=\linewidth]{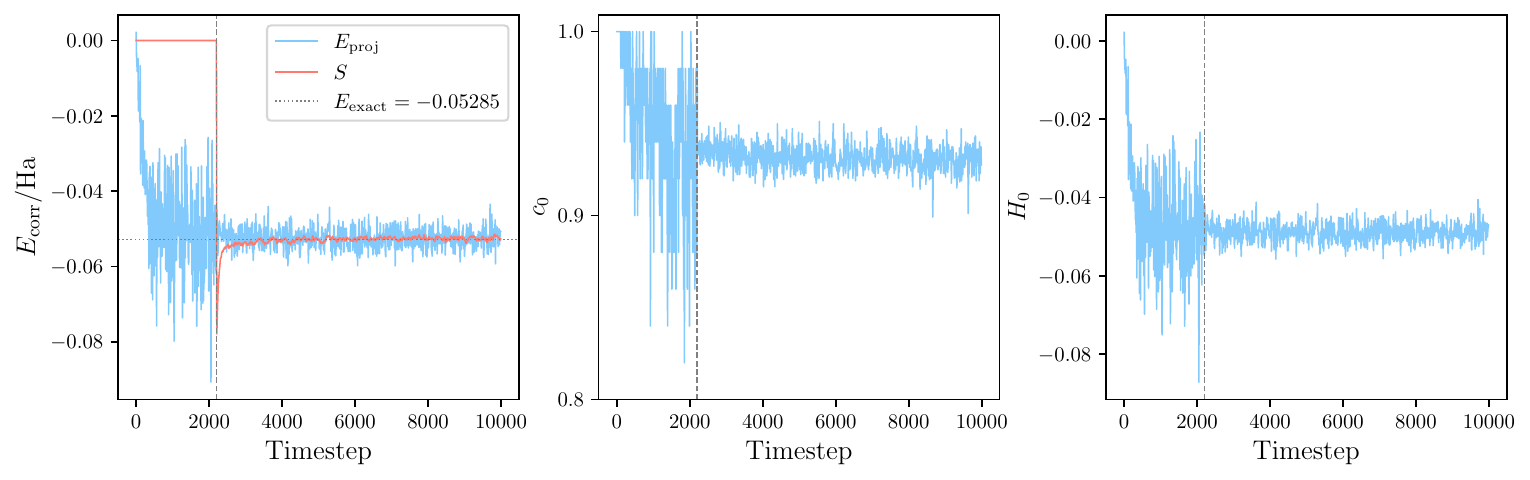}
    \caption{Instantaneous observable values along an MC-PQE trajectory for LiH/STO-3G at $r_\text{LiH} = 2.5$\AA, using the SW shot allocation with $N_s = 10^6$.}
    \label{fig:lih_traj}
\end{figure}

\begin{table}[]
    \centering
    \begin{tabular}{|c|c|c|c|c|c|c|c|}
         \hline\textbf{Observable}&\makecell{\textbf{Variance}\\\textbf{Wavefunction}}& \textbf{SI E} & \textbf{SI PV} & \textbf{SI SW} & \textbf{SI EV} & \textbf{SI ED-PV} & \textbf{SI ED-SW}\\
         \hline
        \multirow{3}{*}{${S}$}&HF & \multirow{3}{*}{0.56}&0.64&0.57&0.78&0.58&0.52\\
        &HF + QMC &&0.54&\textbf{0.47}&0.56&0.50&0.61 \\
        &FCI &&0.53&0.47&0.47&0.62&0.56\\
        \hline
        \multirow{3}{*}{${E_\text{proj}}$}&HF & \multirow{3}{*}{1.58}&2.01&2.00&2.03&1.85&1.72\\
        &HF + QMC &&1.31&1.32&1.40&1.26&\textbf{1.02} \\
        &FCI &&1.44&1.26&1.27&1.17&1.01\\
        \hline
    \end{tabular}
    \caption{Standard error in mHartree of observables computed from MC-PQE runs with different measurement distributions for H$_4$/STO-3G, with $r_\text{HH}=1.5${\AA}   and $N_\text{shots}=10^5$.}
    \label{tab:h4_2}
\end{table}
 \begin{figure}[h!]
     \centering
     \includegraphics[width=\linewidth]{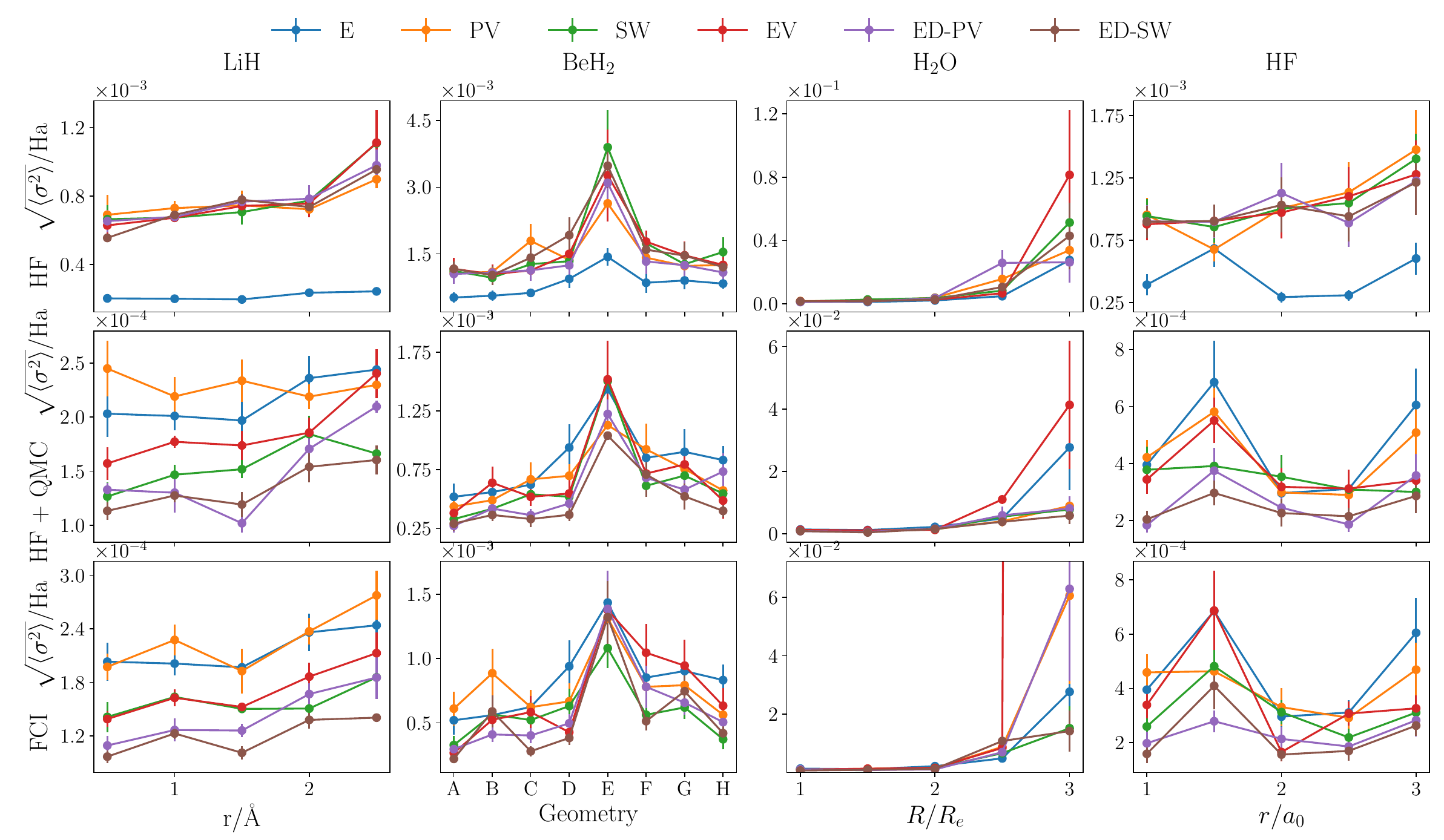}
     \caption{Average standard error in the projected energy obtained by MC-PQE, using different measurement allocation techniques and estimated variance sources. The number of samples is $N_s = 10^6$ for LiH, BeH$_2$ and H$_2$O and $N_s = 10^7$ for HF. Error bars are obtained from 5 independent simulations. The even allocation (E) is independent of variance source.}
     \label{fig:vars_eproj}
 \end{figure}
 We explore the performance of different shot allocation methods for four different first row hydrides (LiH, BeH$_2$, H$_2$O and HF), with Hamiltonians described in \Cref{tab:grouping}. In all cases the systems are treated in the STO-3G basis set, with one frozen core orbital. For LiH, we consider $r_\text{LiH}/$\AA\ \ $\in \{0.5, 1.0, 1.5, 2.0, 2.5\}$. For BeH$_2$, we use the model geometries defined in Ref. \citenum{Purvis1983}, while HF and H$_2$O match Ref. \citenum{Das2010}. The quality of the different variance schemes for each system are shown in \Cref{fig:vars_eproj,fig:vars_shift}, for the projected energy and shift respectively. We note that the shift is more resilient to the variance source, while for the projected energy we observe the expected improvement in quality going from HF variance to correlated wavefunctions.
  The HF + QMC variance scheme shows a marked improvement, which can be justified by considering an example MC-PQE trajectory (see \Cref{fig:lih_traj}). When the shift is turned on, the shot allocation is recomputed, leading to significantly decreased noise in both the instantaneous projected energy and the $c_0$ and $H_0$ estimators used to obtain the final average $E_\text{proj}$. For LiH, BeH$_2$ and HF, the HF + QMC results are very similar to those obtained from the FCI wavefunction, justified again by the similarity between the two. For H$_2$O, we see an increasing discrepancy as the bond length increases and UCCSD becomes a poorer approximation for FCI. In this case, it is unsurprisingly better to use the variance from the wavefunction present in the calculation than an (expensive) FCI oracle.

The optimal shot allocation scheme is different between the two estimators. For $E_\text{proj}$, ED-PV and ED-SW are uniformly best, with pure PV or SW competitive at particular geometries. $H_0$ and $c_0$ generally have much smaller variances than the other terms, particularly near equilibrium geometries, and therefore benefit from methods that do not allocate purely based on variance. This translates to lower variances in the final $E_\text{proj}$ estimator. For $S$, the difference is once again less pronounced, but shift-weighted estimators (SW and ED-SW) tend to outperform methods that do not account for it.

\subsection{Shadow tomography}

\begin{figure}
    \centering
    \includegraphics[width=\linewidth]{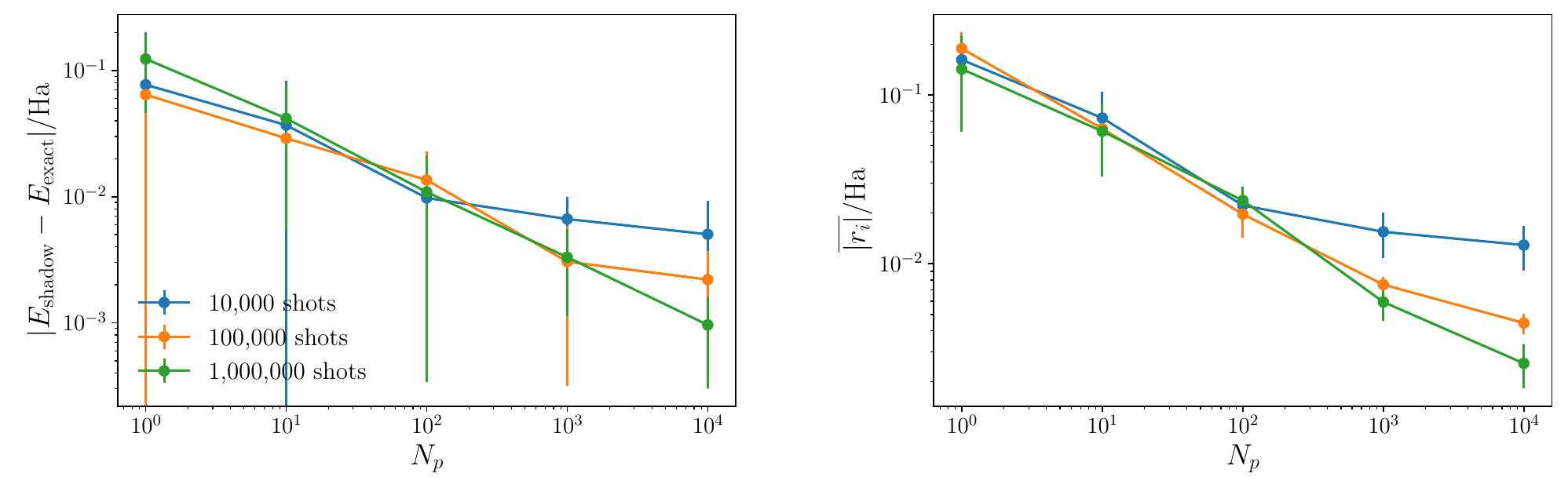}
    \caption{Error in the energy and residuals estimated using shadow tomography for H$_4$/STO-3G at $r_\text{HH} = 1.5$\AA for the exact FCI ground state. All residuals should be zero. Error bars are obtained from five independent trials.}
    \label{fig:H4_shadows_shots}
\end{figure}

We explore the quality of shift and projected energy estimators obtained from a partial shadow tomography of the MC-PQE state. In practice, generating and manipulating the Clifford unitaries is one of the bottlenecks in this algorithm, which would remain even in a true quantum simulation, as the approach requires the classical computation of the expectation values in \Cref{eq:c_exp}. In order to limit this cost, we consider decreasing the number of Clifford gates sampled. Rather than generating an independent Clifford for each shot, we construct a pool of size $N_p$ and sample each one $N_s/N_p$ times. We compare these approaches to estimate the ground state energy and residuals of H$_4$, with results given in \Cref{fig:H4_shadows_shots}.  We find that both energy and residual estimation are relatively robust to the total number of shots, showing very similar behaviour up to large numbers of sampled Clifford matrices. At moderate numbers of selected Cliffords, accuracy is largely independent of total shot count. As residuals are not variationally bounded quantities, we expect errors may change sign over the course of a simulation and lead to favourable error cancellation. In order to keep simulation costs tractable, we therefore attempt MC-PQE simulations with relatively small numbers of sampled Clifford unitaries.  

For an MC-PQE calculation, the pool may be generated once before the Monte Carlo simulation, or regenerated at every iteration. Energy estimators from both approaches and their variances for H$_4$/STO-3g are shown in \Cref{tab:h4_shadows}.

\begin{table}[]
    \centering
    \begin{tabular}{|c|c|c|c|c|}
    \hline
      \textbf{ Bondlength}&\textbf{Estimator}&$N_p$  & \textbf{Fixed} & \textbf{Variable}\\
       \hline
        \multirow{6}{*}{1.0}&\multirow{3}{*}{$E_\text{proj}$}&50 & -0.06826(3) & -0.067(1))\\
        &&100 &-0.0692(6) & -0.0674(6)  \\
        &&200 & -0.06749(4) & -0.0682(7) \\
        \cline{2-5}
        &\multirow{3}{*}{$S$}&50 &  -0.06817(9)& -0.068(1) \\
        &&100 & -0.06782(3) & -0.0681(8)\\
        &&200 & -0.06757(5) & -0.0677(7) \\
        \hline
        \multirow{6}{*}{1.5}&\multirow{3}{*}{$E_\text{proj}$}&50 & -0.16368(5)& -0.172(3)\\
        &&100&-0.17466(5)& -0.170(3)\\
        &&200&-0.17167(5) & -0.169(1)\\
        \cline{2-5}
        &\multirow{3}{*}{$S$}&50 &-0.16379(4)& -0.1687(5)\\
        &&100 &-0.17461(3)  & -0.1680(7)\\
        &&200 &-0.17165(3)& -0.1678(4)\\
        \hline
    \end{tabular}
    \caption{Shift and projected energy estimators from MC-PQE calculations on H$_4$ using partial shadow tomography at different bond, with $N_s=10^6$.}
    \label{tab:h4_shadows}
\end{table}
We observe two trends: fixing the Clifford unitaries before the QMC propagation leads to much less noisy estimators. However, they display a clear bias relative to the exact answer, which grows as the bond-length increases. For $r=1.5$\AA, this approach becomes noticeably non-variational ($E_\text{FCI}(r=1.0$\AA$) = -0.06784$; $E_\text{FCI}(r=1.5$\AA$) = -0.16701$). The bias does not follow a clear pattern with increasing $N_p$ and we suspect it is highly dependent on the exact pool chosen in a given simulation. Therefore, while it is the most cost-effective approach, the fixed Clifford estimator is not reliable. When considering the variable Clifford estimator, we find that, for the numbers of independent Clifford unitaries reported here, it slightly underperforms relative to the shot-based methods in \Cref{tab:h4_2}, and variances decrease slowly with increasing numbers of sampled Cliffords. We observe similar patterns in LiH (\Cref{tab:lih_shadows}), although due to the smaller correlation energies in this system ($E_\text{FCI}(r=1.5$\AA$) =-0.01900467$,
$E_\text{FCI}(r=2.0$\AA$) =-0.03018219$), the biased fixed-pool estimator is within chemical accuracy of the exact energy. Standard errors for the variable-pool estimator remain larger than the comparable results for direct Hamiltonian sampling. Therefore we conclude that for systems of this size, direct Hamiltonian measurement is preferable.  We expect to see a cross-over in performance for larger system sizes, due to the logarithmic scaling of the shadow tomography approach. However computational resources limit the feasibility of this study for large enough systems, using the available Clifford simulators in Qiskit.\cite{JavadiAbhari2024} 
\begin{table}[]
    \centering
    \begin{tabular}{|c|c|c|c|c|}
    \hline
      \textbf{Bondlength}&\textbf{Estimator}&$N_p$  & \textbf{Fixed} & \textbf{Variable}\\
       \hline
        \multirow{4}{*}{1.5}&\multirow{2}{*}{$E_\text{proj}$}&50 & -0.01877(2) & -0.0191(4) \\
        &&100 & -0.01878(2) & -0.0190(4) \\
        \cline{2-5}
        &\multirow{2}{*}{$S$}&50 & -0.01877(2) & -0.0186(7) \\
        &&100 & -0.01880(1) & -0.0190(3) \\
        \hline
        \multirow{4}{*}{2.0}&\multirow{2}{*}{$E_\text{proj}$}&50 & -0.02994(1) & -0.0303(6) \\
        &&100 & -0.02990(2) & -0.0299(5) \\
        \cline{2-5}
        &\multirow{2}{*}{$S$}&50 & -0.02994(3) & -0.0304(6) \\
        &&100 & -0.02992(2) & -0.0297(2) \\
        \hline
    \end{tabular}
    \caption{Shift and projected energy estimators from MC-PQE calculations on LiH using partial shadow tomography at different bond lengths, with $N_s=10^6$.}
    \label{tab:lih_shadows}
\end{table}

\section{Conclusions}\label{sec:conclusions}

In this paper, we have extended efficient observable measurement and estimation techniques developed for VQE to the measurement of the $H_i$ and $c_i$ terms needed in MC-PQE residuals.

We have implemented a range of measurement allocation techniques for the MC-PQE residuals, with the aim of minimising the variance of the Monte Carlo shift and projected energy estimators. We find that the Hartree--Fock wavefunction does not provide a sufficiently good estimate of the variance for measurement distribution, but the instantaneous Monte Carlo wavefunction is very well suited for this purpose and we have easy access to it over the course of the simulation. We therefore do not require a separate heuristic for the variance, like that employed in Ref. \citenum{Yen2023}.

We find that full commuting groups generated by sorted insertion reduce measurement overhead significantly compared to QWC, as expected, with measurement allocation usually leading to a further $2$-$3\times$ reduction in standard error for a given measurement count. In general, distributions that take into account the magnitude of the shift out-perform those that do not, with the ED-SW (even split between different residuals, shift-weighted within a residual) approach providing a good compromise for both $S$ and $E_\text{proj}$ for all systems considered.

We also considered using Lenihan \textit{et al.}'s excitation amplitude sampling through classical shadows technique\cite{Lenihan2025} to estimate the residuals. However, we have found that MC-PQE requires us to resample the Clifford group at every step to obtain an unbiased estimator, which makes simulations of this technique expensive with general Clifford simulators like those available in Qiskit. Larger operator pools would be required to obtain satisfactory accuracy from this approach, although we expect it to become more competitive as system size increases. Developing specific Clifford simulation software is required to further explore this hypothesis.

\begin{acknowledgement}
MAF acknowledges financial support for this work from Peterhouse, Cambridge through a Research Fellowship, and from Downing College, Cambridge through the Kim and Julianna Silverman Research Fellowship. The authors thank George Booth for helpful discussions.
\end{acknowledgement}

\bibliography{qc_qmc}

@article{Arrazola2022,
  doi = {10.22331/q-2022-06-20-742},
  url = {https://doi.org/10.22331/q-2022-06-20-742},
  title = {Universal quantum circuits for quantum chemistry},
  author = {Arrazola, Juan Miguel and Di Matteo, Olivia and Quesada, Nicol{\'{a}}s and Jahangiri, Soran and Delgado, Alain and Killoran, Nathan},
  journal = {{Quantum}},
  issn = {2521-327X},
  publisher = {{Verein zur F{\"{o}}rderung des Open Access Publizierens in den Quantenwissenschaften}},
  volume = {6},
  pages = {742},
  month = jun,
  year = {2022}
}

@article{Bartlett1989,
author = {Bartlett, Rodney J. and Kucharski, Stanislaw A. and Noga, Jozef},
issn = {00092614},
journal = {Chem. Phys. Lett.},
month = {feb},
number = {1},
pages = {133-140},
title = {{Alternative coupled-cluster ans{\"{a}}tze II. The unitary coupled-cluster method}},
url = {https://linkinghub.elsevier.com/retrieve/pii/S0009261489873725},
volume = {155},
year = {1989}
}

@article{Booth2009,
author = {Booth, George H and Thom, Alex J W and Alavi, Ali},
doi = {10.1063/1.3193710},
isbn = {0021-9606},
issn = {00219606},
journal = {J. Chem. Phys.},
number = {5},
pmid = {19673550},
title = {{Fermion Monte Carlo without fixed nodes: A game of life, death, and annihilation in Slater determinant space}},
volume = {131},
year = {2009},
pages = {054106}
}

@article{Brand2022,
  title = {Stochastic differential equation approach to understanding the population control bias in full configuration interaction quantum Monte Carlo},
  author = {Brand, Joachim and Yang, Mingrui and Pahl, Elke},
  journal = {Phys. Rev. B},
  volume = {105},
  issue = {23},
  pages = {235144},
  numpages = {23},
  year = {2022},
  month = {Jun},
  publisher = {American Physical Society},
  doi = {10.1103/PhysRevB.105.235144},
  url = {https://link.aps.org/doi/10.1103/PhysRevB.105.235144}
}

@article{Cerezo2021,
  author  = {Cerezo, M. and Sone, Akira and Volkoff, Tyler and Cincio, Lukasz and Coles, Patrick J.},
  title   = {Cost function dependent barren plateaus in shallow parametrized quantum circuits},
  journal = {Nat. Commun.},
  year    = {2021},
  volume  = {12},
  number  = {1},
  pages   = {1791},
  doi     = {10.1038/s41467-021-21728-w},
  url     = {https://doi.org/10.1038/s41467-021-21728-w}
}

@article{Choy2025,
author = {{Choy B.} and Filip, Maria-Andreea and Wales, David J.},
title = {Energy Landscapes for the Unitary Coupled Cluster Ansatz},
journal = {J. Chem. Theory Comput.},
volume = {21},
number = {4},
pages = {1739-1751},
year = {2025},
doi = {10.1021/acs.jctc.4c01667},
}

@article{Das2010,
    author = {Das, Sanghamitra and Mukherjee, Debashis and Kállay, Mihály},
    title = "{Full implementation and benchmark studies of Mukherjee’s state-specific multireference coupled-cluster ansatz}",
    journal = {J. Chem. Phys.},
    volume = {132},
    number = {7},
    year = {2010},
    month = {02},
    issn = {0021-9606},
    pages = 074103,
    url = {https://doi.org/10.1063/1.3310288},
}

@article{Feynman1982,
author = {Feynman, Richard P},
title = {Simulating physics with computers},
year = {1982},
journal = {Int. J. Theor. Phys.},
volume = {21},
pages = {467-488}
}

@article{Filip2020,
    author = {Filip, Maria-Andreea and Thom, Alex J. W.},
    title = "{A stochastic approach to unitary coupled cluster}",
    journal = {J. Chem. Phys.},
    volume = {153},
    number = {21},
    year = {2020},
    month = {12},
    issn = {0021-9606},
    pages = {214106},
    doi = {10.1063/5.0026141},
    url = {https://doi.org/10.1063/5.0026141},
    
}

@article{Filip2024,
author = {Filip, Maria-Andreea},
title = {Fighting Noise with Noise: A Stochastic Projective Quantum Eigensolver},
journal = {J. Chem. Theory Comput.},
volume = {20},
number = {14},
pages = {5964-5981},
year = {2024},
doi = {10.1021/acs.jctc.4c00295},}

@article{Flyvbjerg1989,
    author = {Flyvbjerg, H. and Petersen, H. G.},
    title = "{Error estimates on averages of correlated data}",
    journal = {J. Chem. Phys.},
    volume = {91},
    number = {1},
    pages = {461-466},
    year = {1989},
    month = {07},
    issn = {0021-9606},
    doi = {10.1063/1.457480},
    url = {https://doi.org/10.1063/1.457480}
}

@ARTICLE{Gokhale2020,
  author={Gokhale, Pranav and Angiuli, Olivia and Ding, Yongshan and Gui, Kaiwen and Tomesh, Teague and Suchara, Martin and Martonosi, Margaret and Chong, Frederic T.},
  journal={IEEE Transa. Quantum Eng.}, 
  title={$O(N^3)$ Measurement Cost for Variational Quantum Eigensolver on Molecular Hamiltonians}, 
  year={2020},
  volume={1},
  number={},
  pages={1-24},
  doi={10.1109/TQE.2020.3035814}}

@article{Grimsley2019b,
author = {Grimsley, Harper R. and Economou, Sophia E. and Barnes, Edwin and Mayhall, Nicholas J.},
doi = {10.1038/s41467-019-10988-2},
issn = {20411723},
journal = {Nat. Commun.},
month = {dec},
number = {1},
pages = {3007},
pmid = {31285433},
publisher = {Nature Publishing Group},
title = {{An adaptive variational algorithm for exact molecular simulations on a quantum computer}},
url = {https://doi.org/10.1038/s41467-019-10988-2},
volume = {10},
year = {2019}
}

@article{Hempel2018,
  title = {Quantum Chemistry Calculations on a Trapped-Ion Quantum Simulator},
  author = {Hempel, Cornelius and Maier, Christine and Romero, Jonathan and McClean, Jarrod and Monz, Thomas and Shen, Heng and Jurcevic, Petar and Lanyon, Ben P. and Love, Peter and Babbush, Ryan and Aspuru-Guzik, Al\'an and Blatt, Rainer and Roos, Christian F.},
  journal = {Phys. Rev. X},
  volume = {8},
  issue = {3},
  pages = {031022},
  numpages = {22},
  year = {2018},
  month = {Jul},
  publisher = {American Physical Society},
  doi = {10.1103/PhysRevX.8.031022},
  url = {https://link.aps.org/doi/10.1103/PhysRevX.8.031022}
}

@article{Kandala2017,
author = {Kandala, Abhinav and Mezzacapo, Antonio and Temme, Kristan and Takita, Maika and Brink, Markus and Chow, Jerry M. and Gambetta, Jay M.},
doi = {10.1038/nature23879},
issn = {14764687},
journal = {Nature},
month = {sep},
number = {7671},
pages = {242-246},
pmid = {28905916},
publisher = {Nature Publishing Group},
title = {{Hardware-efficient variational quantum eigensolver for small molecules and quantum magnets}},
url = {https://www.nature.com/articles/nature23879},
volume = {549},
year = {2017}
}

@article{Kanno2026,
  title = {Quantum-selected configuration interaction: Classical diagonalization of Hamiltonians in subspaces selected by quantum computers},
  author = {Kanno, Keita and Kohda, Masaya and Imai, Ryosuke and Koh, Sho and Mitarai, Kosuke and Mizukami, Wataru and Nakagawa, Yuya O.},
  journal = {Phys. Rev. Res.},
  volume = {8},
  issue = {2},
  pages = {023268},
  numpages = {27},
  year = {2026},
  month = {Jun},
  publisher = {American Physical Society},
  doi = {10.1103/dmn4-snfx},
  url = {https://link.aps.org/doi/10.1103/dmn4-snfx}
}

@article{Kutzelnigg1982,
author = {Kutzelnigg, Werner},
doi = {10.1063/1.444231},
issn = {00219606},
journal = {J. Chem. Phys.},
month = {sep},
number = {6},
pages = {3081-3097},
publisher = {American Institute of Physics},
title = {{Quantum chemistry in Fock space. I. The universal wave and energy operators}},
url = {http://aip.scitation.org/doi/10.1063/1.444231},
volume = {77},
year = {1982}
}

@article{Kutzelnigg1983,
author = {Kutzelnigg, Werner and Koch, Sigurd},
doi = {10.1063/1.446313},
issn = {00219606},
journal = {J. Chem. Phys.},
month = {nov},
number = {9},
pages = {4315-4335},
publisher = {American Institute of Physics},
title = {{Quantum chemistry in Fock space. II. Effective Hamiltonians in Fock space}},
url = {http://aip.scitation.org/doi/10.1063/1.446313},
volume = {79},
year = {1983}
}

@article{Kutzelnigg1984,
author = {Kutzelnigg, Werner},
doi = {10.1063/1.446736},
issn = {00219606},
journal = {J. Chem. Phys.},
month = {jan},
number = {2},
pages = {822-830},
publisher = {American Institute of PhysicsAIP},
title = {{Quantum chemistry in Fock space. III. Particle-hole formalism}},
url = {http://aip.scitation.org/doi/10.1063/1.446736},
volume = {80},
year = {1984}
}

@misc{Lenihan2025,
      title={Excitation Amplitude Sampling for Low Variance Electronic Structure on Quantum Computers}, 
      author={Connor Lenihan and Oliver J. Backhouse and Basil Ibrahim and Tom W. A. Montgomery and Phalgun Lolur and M. J. Bhaseen and George H. Booth},
      year={2025},
      eprint={2506.15438},
      archivePrefix={arXiv},
      primaryClass={quant-ph},
      url={https://arxiv.org/abs/2506.15438}, 
}

@article{McClean2018,
author = {McClean, Jarrod R and Boixo, Sergio and Smelyanskiy, Vadim N and Babbush, Ryan and Neven, Harmut},
journal = {Nat. Commun.},
volume = 9,
pages = 4812,
year = 2018,
title = {Barren plateaus in quantum neural network training landscapes},
url = {https://www.nature.com/articles/s41467-018-07090-4}
}

@article{Peruzzo2014,
  title={A variational eigenvalue solver on a photonic quantum processor},
  author={Peruzzo, Alberto and McClean, Jarrod and Shadbolt, Peter and Yung, Man-Hong and Zhou, Xiao-Qi and Love, Peter J and Aspuru-Guzik, Al{\'a}n and O’brien, Jeremy L},
  journal={Nat. Commun.},
  volume={5},
  number={1},
  pages={4213},
  year={2014},
  publisher={Nature Publishing Group},
  url = {https://www.nature.com/articles/ncomms5213}
}

@misc{Piccinelli2026,
      title={Quantum chemistry with provable convergence via randomized sample-based Krylov quantum diagonalization}, 
      author={Samuele Piccinelli and Alberto Baiardi and Stefano Barison and Max Rossmannek and Almudena Carrera Vazquez and Francesco Tacchino and Stefano Mensa and Edoardo Altamura and Ali Alavi and Mario Motta and Javier Robledo-Moreno and William Kirby and Kunal Sharma and Antonio Mezzacapo and Ivano Tavernelli},
      year={2026},
      eprint={2508.02578},
      archivePrefix={arXiv},
      primaryClass={quant-ph},
      url={https://arxiv.org/abs/2508.02578}, 
}

@article{Preskill2018,
  doi = {10.22331/q-2018-08-06-79},
  url = {https://doi.org/10.22331/q-2018-08-06-79},
  title = {Quantum {C}omputing in the {NISQ} era and beyond},
  author = {Preskill, John},
  journal = {{Quantum}},
  issn = {2521-327X},
  publisher = {{Verein zur F{\"{o}}rderung des Open Access Publizierens in den Quantenwissenschaften}},
  volume = {2},
  pages = {79},
  month = aug,
  year = {2018}
}

@article{Purvis1983,
author = {Purvis III, George D. and Shepard, Ron and Brown, Franklin B. and Bartlett, Rodney J.},
title = {C2V Insertion pathway for BeH2: A test problem for the coupled-cluster single and double excitation model},
journal = {Int. J. Quantum Chem.},
volume = {23},
number = {3},
pages = {835-845},
doi = {https://doi.org/10.1002/qua.560230307},
url = {https://onlinelibrary.wiley.com/doi/abs/10.1002/qua.560230307},
year = {1983}
}

@article{Rajan2025,
    author = {Rajan, Nirmal
Mammavalappil and Khandelwal, Ankit and Nambiar, Manoj and Yamijala, Sharma S. R. K. C.},
    title = {Parallelized Givens
Ansatz for Molecular Ground States:
Bridging Accuracy and Efficiency on NISQ Platforms},
    journal = {J. Phys. Chem. A},
    volume = {129},
    number = {46},
    pages = {10794-10805},
    year = {2025},
    month = {11},
    issn = {1089-5639},
    doi = {10.1021/acs.jpca.5c04605},
    url = {https://doi.org/10.1021/acs.jpca.5c04605},
    eprint = {https://pubs.acs.org/jpcafh/article-pdf/129/46/10794/40789502/jp5c04605.pdf},
}

@article{Reinholdt2025,
  author  = {Reinholdt, Peter and Ziems, Karl Michael and Kjellgren, Erik Rosendahl and Coriani, Sonia and Sauer, Stephan P. A. and Kongsted, Jacob},
  title   = {Critical Limitations in Quantum-Selected Configuration Interaction Methods},
  journal = {J. Chem. Theory Comput.},
  year    = {2025},
  volume  = {21},
  number  = {14},
  pages   = {6811--6822},
  doi     = {10.1021/acs.jctc.5c00375},
  url     = {https://doi.org/10.1021/acs.jctc.5c00375}
}

@article{Reynolds1990,
  author  = {Reynolds, Peter J. and Tobochnik, Jan and Gould, Harvey},
  title   = {Diffusion Quantum Monte Carlo},
  journal = {Computers in Physics},
  year    = {1990},
  volume  = {4},
  number  = {6},
  pages   = {662--668},
  doi     = {10.1063/1.4822960},
  url     = {https://doi.org/10.1063/1.4822960}
}

@article{Shajan2025,
  author  = {Shajan, Akhil and Kaliakin, Danil and Mitra, Abhishek and Robledo Moreno, Javier and Li, Zhen and Motta, Mario and Johnson, Caleb and Saki, Abdullah Ash and Das, Susanta and Sitdikov, Iskandar and Mezzacapo, Antonio and Merz, Kenneth M.},
  title   = {Toward Quantum-Centric Simulations of Extended Molecules: Sample-Based Quantum Diagonalization Enhanced with Density Matrix Embedding Theory},
  journal = {J. Chem. Theory Comput.},
  year    = {2025},
  volume  = {21},
  number  = {14},
  pages   = {6801--6810},
  doi     = {10.1021/acs.jctc.5c00114},
  url     = {https://doi.org/10.1021/acs.jctc.5c00114}
}

@article{Shea2017,
author = {Shea, Jacqueline A. R. and Neuscamman, Eric},
title = {Size Consistent Excited States via Algorithmic Transformations between Variational Principles},
journal = {J. Chem. Theory Comput.},
volume = {13},
number = {12},
pages = {6078-6088},
year = {2017},
doi = {10.1021/acs.jctc.7b00923},
URL = { 
        https://doi.org/10.1021/acs.jctc.7b00923
    
},
}

@article{Sivarajah2021,
doi = {10.1088/2058-9565/ab8e92},
url = {https://dx.doi.org/10.1088/2058-9565/ab8e92},
year = {2020},
month = {nov},
publisher = {IOP Publishing},
volume = {6},
number = {1},
pages = {014003},
author = {Seyon Sivarajah and Silas Dilkes and Alexander Cowtan and Will Simmons and Alec Edgington and Ross Duncan},
title = {t|ket⟩: a retargetable compiler for NISQ devices},
journal = {Quantum Sci. Tech.},
}

@article{Stair2021,
  title = {Simulating Many-Body Systems with a Projective Quantum Eigensolver},
  author = {Stair, Nicholas H. and Evangelista, Francesco A.},
  journal = {PRX Quantum},
  volume = {2},
  issue = {3},
  pages = {030301},
  numpages = {20},
  year = {2021},
  month = {Jul},
  publisher = {American Physical Society},
  doi = {10.1103/PRXQuantum.2.030301},
  url = {https://link.aps.org/doi/10.1103/PRXQuantum.2.030301}
}

@article{Thom2010,
author = {Thom, Alex J W},
doi = {10.1103/PhysRevLett.105.263004},
issn = {00319007},
journal = {Phys. Rev. Lett.},
number = {26},
pages = {263004},
pmid = {21231654},
title = {{Stochastic coupled cluster theory}},
volume = {105},
year = {2010}
}

@article{Wang2021,
  author  = {Wang, Samson and Fontana, Enrico and Cerezo, M. and Sharma, Kunal and Sone, Akira and Cincio, Lukasz and Coles, Patrick J.},
  title   = {Noise-induced barren plateaus in variational quantum algorithms},
  journal = {Nat. Commun.},
  year    = {2021},
  volume  = {12},
  number  = {1},
  pages   = {6961},
  doi     = {10.1038/s41467-021-27045-6},
  url     = {https://doi.org/10.1038/s41467-021-27045-6}
}

@misc{Welling2026,
      title={Shallow Electronic State Preparation for Quantum Chemistry with Quantum Monte Carlo Pre-Selection}, 
      author={Eline Welling and Lila Cadi-Tazi and Alex J. W. Thom and Maria-Andreea Filip},
      year={2026},
      eprint={2605.31139},
      archivePrefix={arXiv},
      primaryClass={quant-ph},
      url={https://arxiv.org/abs/2605.31139}, 
}

@article{Yang2020,
    author = {Yang, Mingrui and Pahl, Elke and Brand, Joachim},
    title = "{Improved walker population control for full configuration interaction quantum Monte Carlo}",
    journal = {J. Chem. Phys},
    volume = {153},
    number = {17},
    pages = {174103},
    year = {2020},
    month = {11},
    issn = {0021-9606},
    doi = {10.1063/5.0023088},
}

@article{Yen2023,
author = {Yen, T C. and Ganeshram, A. and Izmaylov, A. F},
title = {Deterministic improvements of quantum measurements with grouping of compatible operators, non-local transformations, and covariance estimates},
journal = {npj Quantum Inf},
volume = 9,
pages = 14,
year = 2023,
doi={https://doi.org/10.1038/s41534-023-00683-y}}

@article{Choi2023,
  doi = {10.22331/q-2023-01-03-889},
  url = {https://doi.org/10.22331/q-2023-01-03-889},
  title = {Fluid fermionic fragments for optimizing quantum measurements of electronic {H}amiltonians in the variational quantum eigensolver},
  author = {Choi, Seonghoon and Loaiza, Ignacio and Izmaylov, Artur F.},
  journal = {{Quantum}},
  issn = {2521-327X},
  publisher = {{Verein zur F{\"{o}}rderung des Open Access Publizierens in den Quantenwissenschaften}},
  volume = {7},
  pages = {889},
  month = jan,
  year = {2023}
}

@article{Choi2022,
author = {Choi, Seonghoon and Yen, Tzu-Ching and Izmaylov, Artur F.},
title = {Improving Quantum Measurements by Introducing “Ghost” Pauli Products},
journal = {J. Chem. Theory Comput.},
volume = {18},
number = {12},
pages = {7394-7402},
year = {2022},
doi = {10.1021/acs.jctc.2c00837},

URL = { 
    
        https://doi.org/10.1021/acs.jctc.2c00837
    
    

},

}

@article{Huang2021,
  title = {Efficient Estimation of Pauli Observables by Derandomization},
  author = {Huang, Hsin-Yuan and Kueng, Richard and Preskill, John},
  journal = {Phys. Rev. Lett.},
  volume = {127},
  issue = {3},
  pages = {030503},
  numpages = {7},
  year = {2021},
  month = {Jul},
  publisher = {American Physical Society},
  doi = {10.1103/PhysRevLett.127.030503},
  url = {https://link.aps.org/doi/10.1103/PhysRevLett.127.030503}
}

@article{Wu2023,
  doi = {10.22331/q-2023-01-13-896},
  url = {https://doi.org/10.22331/q-2023-01-13-896},
  title = {Overlapped grouping measurement: {A} unified framework for measuring quantum states},
  author = {Wu, Bujiao and Sun, Jinzhao and Huang, Qi and Yuan, Xiao},
  journal = {{Quantum}},
  issn = {2521-327X},
  publisher = {{Verein zur F{\"{o}}rderung des Open Access Publizierens in den Quantenwissenschaften}},
  volume = {7},
  pages = {896},
  month = jan,
  year = {2023}
}

@article{Huang2020,
    author = {Huang, H Y. and Kueng, R. and Preskill, J.},
    title = {Predicting many properties of a quantum system from very few measurements},
    journal = {Nat. Phys.},
    year = 2020,
    volume = 16,
    pages = {1050–1057},
    doi = {https://doi.org/10.1038/s41567-020-0932-7}
}

@article{Misiewicz2024,
author = {Misiewicz, Jonathon P. and Evangelista, Francesco A.},
title = {Implementation of the Projective Quantum Eigensolver on a Quantum Computer},
journal = {J. Phys. Chem. A},
volume = {128},
number = {11},
pages = {2220-2235},
year = {2024},
doi = {10.1021/acs.jpca.3c07429},

URL = { 
    
        https://doi.org/10.1021/acs.jpca.3c07429
    
    

},
eprint = { 
    
        https://doi.org/10.1021/acs.jpca.3c07429
    
    

}

}

@article{Hadfield2022,
   author = {Charles Hadfield and Sergey Bravyi and Rudy Raymond and Antonio Mezzacapo},
   doi = {10.1007/S00220-022-04343-8},
   issn = {1432-0916},
   issue = {3},
   journal = {Comm. Math. Phys.},
   month = {3},
   pages = {951-967},
   publisher = {Springer},
   title = {Measurements of Quantum Hamiltonians with Locally-Biased Classical Shadows},
   volume = {391},
   url = {https://link.springer.com/article/10.1007/s00220-022-04343-8},
   year = {2022}
}

@article{Crawford2021,
   author = {Ophelia Crawford and Barnaby Van Straaten and Daochen Wang and Thomas Parks and Earl Campbell and Stephen Brierley},
   doi = {10.22331/q-2021-01-20-385},
   issn = {2521327X},
   journal = {Quantum},
   month = {1},
   pages = {385},
   publisher = {Verein zur Förderung des Open Access Publizierens in den Quantenwissenschaften},
   title = {Efficient quantum measurement of Pauli operators in the presence of finite sampling error},
   volume = {5},
   url = {https://quantum-journal.org/papers/q-2021-01-20-385/},
   year = {2021}
}

@article{Patel2025,
   author = {Smik Patel and Praveen Jayakumar and Tzu Ching Yen and Artur F. Izmaylov},
   doi = {10.1021/ACS.CHEMREV.5C00055},
   issn = {15206890},
   issue = {16},
   journal = {Chem. Rev.},
   month = {8},
   pages = {7490-7524},
   pmid = {40690271},
   publisher = {American Chemical Society},
   title = {Quantum Measurement for Quantum Chemistry on a Quantum Computer},
   volume = {125},
   url = {/doi/pdf/10.1021/acs.chemrev.5c00055?ref=article_openPDF},
   year = {2025}
}

@article{PsiQuantum2025Manufacturable,
  author = {
    Alexander, Koen and Benyamini, Avishai and Black, Dylan and Bonneau, Damien and
    Burgos, Stanley and Burridge, Ben and Cable, Hugo and Campbell, Geoff and
    Catalano, Gabriel and Ceballos, Alejandro and Chang, Chia-Ming and
    Sen Choudhury, Sourav and Chung, C. J. and Danesh, Fariba and Dauer, Tom and
    Davis, Michael and Dudley, Eric and Er-Xuan, Ping and Fargas, Josep and
    Farsi, Alessandro and Fenrich, Colleen and Frazer, Jonathan and Fukami, Masaya and
    Ganesan, Yogeeswaran and Gibson, Gary and Gimeno-Segovia, Mercedes and
    Goeldi, Sebastian and Goley, Patrick and Haislmaier, Ryan and Halimi, Sami and
    Hansen, Paul and Hardy, Sam and Horng, Jason and House, Matthew and Hu, Hong and
    Jadidi, Mehdi and Jain, Vijay and Johansson, Henrik and Jones, Thomas and
    Kamineni, Vimal and Kelez, Nicholas and Koustuban, Ravi and Kovall, George and
    Krogen, Peter and Kumar, Nikhil and Liang, Yong and LiCausi, Nicholas and
    Llewellyn, Dan and Lokovic, Kimberly and Lovelady, Michael and
    Manfrinato, Vitor Riseti and Melnichuk, Ann and Mendoza, Gabriel and
    Moores, Brad and Mukherjee, Shaunak and Munns, Joseph and
    Musalem, Francois-Xavier and Najafi, Faraz and O'Brien, Jeremy L. and
    Ortmann, J. Elliott and Pai, Sunil and Park, Bryan and Peng, Hsuan-Tung and
    Penthorn, Nicholas and Peterson, Brennan and Peterson, Gabriel and
    Poush, Matt and Pryde, Geoff J. and Ramprasad, Tarun and Ray, Gareth and
    Viejo Rodriguez, Angelita and Roxworthy, Brian and Rudolph, Terry and
    Saunders, Dylan J. and Shadbolt, Pete and Shah, Deesha and
    Bahgat Shehata, Andrea and Shin, Hyungki and Sinsky, Jeffrey and Smith, Jake and
    Sohn, Ben and Sohn, Young-Ik and Son, Gyeongho and Souza, Mario C. M. M. and
    Sparrow, Chris and Staffaroni, Matteo and Stavrakas, Camille and
    Sukumaran, Vijay and Tamborini, Davide and Thompson, Mark G. and
    Tran, Khanh and Triplett, Mark and Tung, Maryann and Veitia, Andrzej and
    Vert, Alexey and Vidrighin, Mihai D. and Vorobeichik, Ilya and Weigel, Peter and
    Wingert, Matthew and Wooding, Jamie and Zhou, Xinran
  },
  title = {A manufacturable platform for photonic quantum computing},
  journal = {Nature},
  volume = {641},
  pages = {876--883},
  year = {2025},
  doi = {10.1038/s41586-025-08820-7}
}

@article{King2025BeyondClassical,
  author = {
    King, Andrew D. and Nocera, Alberto and Rams, Marek M. and Dziarmaga, Jacek and
    Wiersema, Roeland and Bernoudy, William and Raymond, Jack and Kaushal, Nitin and
    Heinsdorf, Niclas and Harris, Richard and Boothby, Kelly and Altomare, Fabio and
    Asad, Mohsen and Berkley, Andrew J. and Boschnak, Martin and Chern, Kevin and
    Christiani, Holly and Cibere, Samantha and Connor, Jake and Dehn, Martin H. and
    Deshpande, Rahul and Ejtemaee, Sara and Farre, Pau and Hamer, Kelsey and
    Hoskinson, Emile and Huang, Shuiyuan and Johnson, Mark W. and Kortas, Samuel and
    Ladizinsky, Eric and Lanting, Trevor and Lai, Tony and Li, Ryan and
    MacDonald, Allison J.R. and Marsden, Gaelen and McGeoch, Catherine C. and
    Molavi, Reza and Oh, Travis and Neufeld, Richard and Norouzpour, Mana and
    Pasvolsky, Joel and Poitras, Patrick and Poulin-Lamarre, Gabriel and
    Prescott, Thomas and Reis, Mauricio and Rich, Chris and Samani, Mohammad and
    Sheldan, Benjamin and Smirnov, Anatoly and Sterpka, Edward and
    Trullas Clavera, Berta and Tsai, Nicholas and Volkmann, Mark and
    Whiticar, Alexander M. and Whittaker, Jed D. and Wilkinson, Warren and
    Yao, Jason and Yi, T. J. and Sandvik, Anders W. and Alvarez, Gonzalo and
    Melko, Roger G. and Carrasquilla, Juan and Franz, Marcel and Amin, Mohammad H.
  },
  title = {Beyond-classical computation in quantum simulation},
  journal = {Science},
  volume = {388},
  number = {6743},
  pages = {199--204},
  year = {2025},
  doi = {10.1126/science.ado6285}
}

@article{Bartee2025SpinQubitCMOS,
  author = {
    Bartee, Samuel K. and Gilbert, Will and Zuo, Kun and Das, Kushal and
    Tanttu, Tuomo and Yang, Chih Hwan and Dumoulin Stuyck, Nard and
    Pauka, Sebastian J. and Su, Rocky Y. and Lim, Wee Han and Serrano, Santiago and
    Escott, Christopher C. and Hudson, Fay E. and Itoh, Kohei M. and Laucht, Arne and
    Dzurak, Andrew S. and Reilly, David J.
  },
  title = {Spin-qubit control with a milli-kelvin CMOS chip},
  journal = {Nature},
  volume = {643},
  pages = {382--387},
  year = {2025},
  doi = {10.1038/s41586-025-09157-x}
}

@article{Steinacker2025IndustryCompatible,
  author = {
    Steinacker, Paul and Dumoulin Stuyck, Nard and Lim, Wee Han and Tanttu, Tuomo and
    Feng, MengKe and Serrano, Santiago and Nickl, Andreas and Candido, Marco and
    Cifuentes, Jesus D. and Vahapoglu, Ensar and Bartee, Samuel K. and
    Hudson, Fay E. and Chan, Kok Wai and Kubicek, Stefan and Jussot, Julien and
    Canvel, Yann and Beyne, Sofie and Shimura, Yosuke and Loo, Roger and
    Godfrin, Clement and Raes, Bart and Baudot, Sylvain and Wan, Danny and
    Laucht, Arne and Yang, Chih Hwan and Saraiva, Andre and Escott, Christopher C. and
    De Greve, Kristiaan and Dzurak, Andrew S.
  },
  title = {Industry-compatible silicon spin-qubit unit cells exceeding 99\% fidelity},
  journal = {Nature},
  volume = {646},
  pages = {81--87},
  year = {2025},
  doi = {10.1038/s41586-025-09531-9}
}

@article{Bluvstein2026FaultTolerant,
  author = {
    Bluvstein, Dolev and Geim, Alexandra A. and Li, Sophie H. and Evered, Simon J. and
    Bonilla Ataides, J. Pablo and Baranes, Gefen and Gu, Andi and Manovitz, Tom and
    Xu, Muqing and Kalinowski, Marcin and Majidy, Shayan and Kokail, Christian and
    Maskara, Nishad and Trapp, Elias C. and Stewart, Luke M. and Hollerith, Simon and
    Zhou, Hengyun and Gullans, Michael J. and Yelin, Susanne F. and Greiner, Markus and
    Vuletić, Vladan and Cain, Madelyn and Lukin, Mikhail D.
  },
  title = {A fault-tolerant neutral-atom architecture for universal quantum computation},
  journal = {Nature},
  volume = {649},
  pages = {39--46},
  year = {2026},
  doi = {10.1038/s41586-025-09848-5}
}

@article{Tan2025,
  title = {Experimental Quantum Error Correction below the Surface Code Threshold via All-Microwave Leakage Suppression},
  author = {He, Tan and Lin, Weiping and Wang, Rui and Li, Yuan and Bei, Jiahao and Cai, Jianbin and Cao, Sirui and Chen, Danning and Chen, Kefu and Chen, Xiawei and Chen, Zhe and Chen, Zhiyuan and Chen, Zihua and Chu, Wenhao and Deng, Hui and Ding, Xun and Ding, Zhuzhengqi and Fan, Bo and Fan, Daojin and Fu, Yuanhao and Gao, Dongxin and Gong, Ming and Gui, Jiacheng and Guo, Cheng and Guo, Shaojun and Han, Lianchen and Hong, Linyin and Hu, Yisen and Huang, He-Liang and Huo, Yong-Heng and Jiang, Chenyan and Jiang, Lei and Jiang, Tao and Jiang, Zuokai and Jin, Honghong and Li, Dayu and Li, Dongdong and Li, Jiaqi and Li, Jinjin and Li, Junyan and Li, Junyun and Li, Na and Li, Shaowei and Li, Yuhuai and Liang, Futian and Liao, Nanxing and Lin, Jin and Liu, Ke and Liu, Maliang and Liu, Yancheng and Lou, Haoxin and Ma, Yuwei and Nan, Kailiang and Nie, Meijuan and Niu, Le and Peng, Wenyi and Qian, Haoran and Rong, Hao and Rong, Tao and Shen, Huiyan and Shen, Qiong and Su, Hong and Su, Feifan and Sun, Chenyin and Sun, Liangchao and Sun, Tianzuo and Sun, Yingxiu and Tan, Yimeng and Tan, Jun and Tu, Wenbing and Wang, Jiafei and Wang, Biao and Wang, Chang and Wang, Chen and Wang, Chu and Wang, Jian and Wang, Shengtao and Wang, Xinzhe and Wei, Zuolin and Wu, Dachao and Wu, Gang and Wu, Yulin and Xu, Yu and Xue, Chun and Yan, Kai and Yan, Xin and Yang, Weifeng and Yang, Xinpeng and Yang, Yang and Ye, Yangsen and Ye, Zhenping and Yi, Zhengzhong and Ying, Chong and Yu, Jiale and Yu, Qinjing and Zeng, Xiangdong and Zha, Chen and Zhan, Shaoyu and Zhang, Haibin and Zhang, He and Zhang, Kaili and Zhang, Wen and Zhang, Yiming and Zhang, Yongzhuo and Zhang, Ziying and Zhao, Guming and Zhao, Xintao and Zhao, Youwei and Zhao, Zhong and Zheng, Luyuan and Zhou, Fei and Zhou, Liang and Zhou, Na and Zhou, Naibin and Zhu, Chengjun and Zhu, Qingling and Zou, Guihong and Zou, Haonan and Zhang, Qiang and Lu, Chao-Yang and Peng, Cheng-Zhi and Chen, Fusheng and Zhu, XiaoBo and Pan, Jian-Wei},
  journal = {Phys. Rev. Lett.},
  volume = {135},
  issue = {26},
  pages = {260601},
  numpages = {7},
  year = {2025},
  month = {Dec},
  publisher = {American Physical Society},
  doi = {10.1103/rqkg-dw31},
  url = {https://link.aps.org/doi/10.1103/rqkg-dw31}
}

@misc{Drouet2026,
      title={Error correction on an array of superconducting qubits with defective components}, 
      author={Julien M. Drouet and Xanda C. Kolesnikow and Campbell K. McLauchlan and Georgia M. Nixon and Seok-Hyung Lee and Dominic J. Williamson and Stephen D. Bartlett and Benjamin J. Brown and Robin Harper},
      year={2026},
      eprint={2607.12118},
      archivePrefix={arXiv},
      primaryClass={quant-ph},
      url={https://arxiv.org/abs/2607.12118}, 
}

@article{Caune2026RealTimeQEC,
  author = {
    Caune, Laura and
    Skoric, Luka and
    Blunt, Nick S. and
    Ruban, Archibald and
    McDaniel, Jimmy and
    Valery, Joseph A. and
    Patterson, Andrew D. and
    Gramolin, Alexander V. and
    Majaniemi, Joonas and
    Barnes, Kenton M. and
    Bialas, Tomasz and
    Buğdaycı, Okan and
    Crawford, Ophelia and
    Gehér, György P. and
    Krovi, Hari and
    Matekole, Elisha and
    Topal, Canberk and
    Poletto, Stefano and
    Bryant, Michael and
    Snyder, Kalan and
    Gillespie, Neil I. and
    Jones, Glenn and
    Johar, Kauser and
    Campbell, Earl T. and
    Hill, Alexander D.
  },
  title = {Demonstrating real-time and low-latency quantum error correction with superconducting qubits},
  journal = {Nat. Commun.},
  volume = {17},
  pages = {7383},
  year = {2026},
  doi = {10.1038/s41467-026-73331-6},
  url = {https://doi.org/10.1038/s41467-026-73331-6}
}

@article{GoogleQuantumAI2025BelowThreshold,
  author = {
    Acharya, Rajeev and
    Abanin, Dmitry A. and
    Aghababaie-Beni, Laleh and
    Aleiner, Igor and
    Andersen, Trond I. and
    Ansmann, Markus and
    Arute, Frank and
    Arya, Kunal and
    Asfaw, Abraham and
    Astrakhantsev, Nikita and
    Atalaya, Juan and
    Babbush, Ryan and
    Bacon, Dave and
    Ballard, Brian and
    Bardin, Joseph C. and
    Bausch, Johannes and
    Bengtsson, Andreas and
    Bilmes, Alexander and
    Blackwell, Sam and
    Boixo, Sergio and
    Bortoli, Gina and
    Bourassa, Alexandre and
    Bovaird, Jenna and
    Brill, Leon and
    Broughton, Michael and
    Browne, David A. and
    Buchea, Brett and
    Buckley, Bob B. and
    Buell, David A. and
    Burger, Tim and
    Burkett, Brian and
    Bushnell, Nicholas and
    Cabrera, Anthony and
    Campero, Juan and
    Chang, Hung-Shen and
    Chen, Yu and
    Chen, Zijun and
    Chiaro, Ben and
    Chik, Desmond and
    Chou, Charina and
    Claes, Jahan and
    Cleland, Agnetta Y. and
    Cogan, Josh and
    Collins, Roberto and
    Conner, Paul and
    Courtney, William and
    Crook, Alexander L. and
    Curtin, Ben and
    Das, Sayan and
    Davies, Alex and
    De Lorenzo, Laura and
    Debroy, Dripto M. and
    Demura, Sean and
    Devoret, Michel and
    Di Paolo, Agustin and
    Donohoe, Paul and
    Drozdov, Ilya and
    Dunsworth, Andrew and
    Earle, Clint and
    Edlich, Thomas and
    Eickbusch, Alec and
    Elbag, Aviv Moshe and
    Elzouka, Mahmoud and
    Erickson, Catherine and
    Faoro, Lara and
    Farhi, Edward and
    Ferreira, Vinicius S. and
    Flores Burgos, Leslie and
    Forati, Ebrahim and
    Fowler, Austin G. and
    Foxen, Brooks and
    Ganjam, Suhas and
    Garcia, Gonzalo and
    Gasca, Robert and
    Genois, Élie and
    Giang, William and
    Gidney, Craig and
    Gilboa, Dar and
    Gosula, Raja and
    Grajales Dau, Alejandro and
    Graumann, Dietrich and
    Greene, Alex and
    Gross, Jonathan A. and
    Habegger, Steve and
    Hall, John and
    Hamilton, Michael C. and
    Hansen, Monica and
    Harrigan, Matthew P. and
    Harrington, Sean D. and
    Heras, Francisco J. H. and
    Heslin, Stephen and
    Heu, Paula and
    Higgott, Oscar and
    Hill, Gordon and
    Hilton, Jeremy and
    Holland, George and
    Hong, Sabrina and
    Huang, Hsin-Yuan and
    Huff, Ashley and
    Huggins, William J. and
    Ioffe, Lev B. and
    Isakov, Sergei V. and
    Iveland, Justin and
    Jeffrey, Evan and
    Jiang, Zhang and
    Jones, Cody and
    Jordan, Stephen and
    Joshi, Chaitali and
    Juhas, Pavol and
    Kafri, Dvir and
    Kang, Hui and
    Karamlou, Amir H. and
    Kechedzhi, Kostyantyn and
    Kelly, Julian and
    Khaire, Trupti and
    Khattar, Tanuj and
    Khezri, Mostafa and
    Kim, Seon and
    Klimov, Paul V. and
    Klots, Andrey R. and
    Kobrin, Bryce and
    Kohli, Pushmeet and
    Korotkov, Alexander N. and
    Kostritsa, Fedor and
    Kothari, Robin and
    Kozlovskii, Borislav and
    Kreikebaum, John Mark and
    Kurilovich, Vladislav D. and
    Lacroix, Nathan and
    Landhuis, David and
    Lange-Dei, Tiano and
    Langley, Brandon W. and
    Laptev, Pavel and
    Lau, Kim-Ming and
    Le Guevel, Loïck and
    Ledford, Justin and
    Lee, Joonho and
    Lee, Kenny and
    Lensky, Yuri D. and
    Leon, Shannon and
    Lester, Brian J. and
    Li, Wing Yan and
    Li, Yin and
    Lill, Alexander T. and
    Liu, Wayne and
    Livingston, William P. and
    Locharla, Aditya and
    Lucero, Erik and
    Lundahl, Daniel and
    Lunt, Aaron and
    Madhuk, Sid and
    Malone, Fionn D. and
    Maloney, Ashley and
    Mandrà, Salvatore and
    Manyika, James and
    Martin, Leigh S. and
    Martin, Orion and
    Martin, Steven and
    Maxfield, Cameron and
    McClean, Jarrod R. and
    McEwen, Matt and
    Meeks, Seneca and
    Megrant, Anthony and
    Mi, Xiao and
    Miao, Kevin C. and
    Mieszala, Amanda and
    Molavi, Reza and
    Molina, Sebastian and
    Montazeri, Shirin and
    Morvan, Alexis and
    Movassagh, Ramis and
    Mruczkiewicz, Wojciech and
    Naaman, Ofer and
    Neeley, Matthew and
    Neill, Charles and
    Neven, Hartmut and
    Newman, Michael and
    Ng, Jiun How and
    Nguyen, Anthony and
    Nguyen, Murray and
    Ni, Chia-Hung and
    Niu, Murphy Yuezhen and
    O'Brien, Thomas E. and
    Oliver, William D. and
    Opremcak, Alex and
    Ottosson, Kristoffer and
    Petukhov, Andre and
    Pizzuto, Alex and
    Platt, John and
    Potter, Rebecca and
    Pritchard, Orion and
    Pryadko, Leonid P. and
    Quintana, Chris and
    Ramachandran, Ganesh and
    Reagor, Matthew J. and
    Redding, John and
    Rhodes, David M. and
    Roberts, Gabrielle and
    Rosenberg, Eliott and
    Rosenfeld, Emma and
    Roushan, Pedram and
    Rubin, Nicholas C. and
    Saei, Negar and
    Sank, Daniel and
    Sankaragomathi, Kannan and
    Satzinger, Kevin J. and
    Schurkus, Henry F. and
    Schuster, Christopher and
    Senior, Andrew W. and
    Shearn, Michael J. and
    Shorter, Aaron and
    Shutty, Noah and
    Shvarts, Vladimir and
    Singh, Shraddha and
    Sivak, Volodymyr and
    Skruzny, Jindra and
    Small, Spencer and
    Smelyanskiy, Vadim and
    Smith, W. Clarke and
    Somma, Rolando D. and
    Springer, Sofia and
    Sterling, George and
    Strain, Doug and
    Suchard, Jordan and
    Szasz, Aaron and
    Sztein, Alex and
    Thor, Douglas and
    Torres, Alfredo and
    Torunbalci, M. Mert and
    Vaishnav, Abeer and
    Vargas, Justin and
    Vdovichev, Sergey and
    Vidal, Guifre and
    Villalonga, Benjamin and
    Vollgraff Heidweiller, Catherine and
    Waltman, Steven and
    Wang, Shannon X. and
    Ware, Brayden and
    Weber, Kate and
    Weidel, Travis and
    White, Theodore and
    Wong, Kristi and
    Woo, Bryan W. K. and
    Xing, Cheng and
    Yao, Z. Jamie and
    Yeh, Ping and
    Ying, Bicheng and
    Yoo, Juhwan and
    Yosri, Noureldin and
    Young, Grayson and
    Zalcman, Adam and
    Zhang, Yaxing and
    Zhu, Ningfeng and
    Zobrist, Nicholas
  },
  title = {Quantum error correction below the surface code threshold},
  journal = {Nature},
  volume = {638},
  number = {8052},
  pages = {920--926},
  year = {2025},
  doi = {10.1038/s41586-024-08449-y},
  url = {https://doi.org/10.1038/s41586-024-08449-y}
}

@article{Yen2020,
    author = {Yen, Tzu-Ching and Verteletskyi, Vladyslav and Izmaylov, Artur F.},
    title = {Measuring All Compatible Operators in One Series of
Single-Qubit Measurements Using Unitary Transformations},
    journal = {J. Chem. Theory Comput.},
    volume = {16},
    number = {4},
    pages = {2400-2409},
    year = {2020},
    month = {03},
    issn = {1549-9618},
    doi = {10.1021/acs.jctc.0c00008},
    url = {https://doi.org/10.1021/acs.jctc.0c00008},
    eprint = {https://pubs.acs.org/jctcce/article-pdf/16/4/2400/18825464/ct0c00008.pdf},
}

@article{Izmaylov2020,
    author = {Izmaylov, Artur F. and Yen, Tzu-Ching and Lang, Robert A. and Verteletskyi, Vladyslav},
    title = {Unitary Partitioning
Approach to the Measurement Problem
in the Variational Quantum Eigensolver Method},
    journal = {J. Chem. Theory Comput.},
    volume = {16},
    number = {1},
    pages = {190-195},
    year = {2019},
    month = {11},
    issn = {1549-9618},
    doi = {10.1021/acs.jctc.9b00791},
    url = {https://doi.org/10.1021/acs.jctc.9b00791},
    eprint = {https://pubs.acs.org/jctcce/article-pdf/16/1/190/10325074/ct9b00791.pdf},
}

@article{Huggins2021,
  author = {
    Huggins, William J. and
    McClean, Jarrod R. and
    Rubin, Nicholas C. and
    Jiang, Zhang and
    Wiebe, Nathan and
    Whaley, K. Birgitta and
    Babbush, Ryan
  },
  title = {Efficient and noise resilient measurements for quantum chemistry on near-term quantum computers},
  journal = {npj Quantum Inf.},
  volume = {7},
  number = {1},
  pages = {23},
  year = {2021},
  month = feb,
  publisher = {Springer Nature},
  doi = {10.1038/s41534-020-00341-7},
  url = {https://doi.org/10.1038/s41534-020-00341-7}
}

@article{Yen2021,
  title = {Cartan Subalgebra Approach to Efficient Measurements of Quantum Observables},
  author = {Yen, Tzu-Ching and Izmaylov, Artur F.},
  journal = {PRX Quantum},
  volume = {2},
  issue = {4},
  pages = {040320},
  numpages = {12},
  year = {2021},
  month = {Oct},
  publisher = {American Physical Society},
  doi = {10.1103/PRXQuantum.2.040320},
  url = {https://link.aps.org/doi/10.1103/PRXQuantum.2.040320}
}

@article{Verteletskyi2020,
    author = {Verteletskyi, Vladyslav and Yen, Tzu-Ching and Izmaylov, Artur F.},
    title = {Measurement optimization in the variational quantum eigensolver using a minimum clique cover},
    journal = {J. Chem. Phys.},
    volume = {152},
    number = {12},
    pages = {124114},
    year = {2020},
    month = {03},
    issn = {0021-9606},
    doi = {10.1063/1.5141458},
    url = {https://doi.org/10.1063/1.5141458},
    eprint = {https://pubs.aip.org/aip/jcp/article-pdf/doi/10.1063/1.5141458/15573883/124114_1_online.pdf},
}

@article{Gena2022,
  title = {Optimization of variational-quantum-eigensolver measurement by partitioning Pauli operators using multiqubit Clifford gates on noisy intermediate-scale quantum hardware},
  author = {Jena, Andrew and Genin, Scott N. and Mosca, Michele},
  journal = {Phys. Rev. A},
  volume = {106},
  issue = {4},
  pages = {042443},
  numpages = {6},
  year = {2022},
  month = {Oct},
  publisher = {American Physical Society},
  doi = {10.1103/PhysRevA.106.042443},
  url = {https://link.aps.org/doi/10.1103/PhysRevA.106.042443}
}

@book{ShumwayStoffer2017,
  author    = {Robert H. Shumway and David S. Stoffer},
  title     = {Time Series Analysis and Its Applications: With R Examples},
  edition   = {4},
  series    = {Springer Texts in Statistics},
  publisher = {Springer Cham},
  year      = {2017},
  doi       = {10.1007/978-3-319-52452-8},
  isbn      = {978-3-319-52452-8}
}

@misc{JavadiAbhari2024,
      title={Quantum computing with Qiskit}, 
      author={Ali Javadi-Abhari and Matthew Treinish and Kevin Krsulich and Christopher J. Wood and Jake Lishman and Julien Gacon and Simon Martiel and Paul D. Nation and Lev S. Bishop and Andrew W. Cross and Blake R. Johnson and Jay M. Gambetta},
      year={2024},
      eprint={2405.08810},
      archivePrefix={arXiv},
      primaryClass={quant-ph},
      url={https://arxiv.org/abs/2405.08810}, 
}

@article{Kurita2023,
    author = {Kurita, Tomochika and Morita, Mikio and Oshima, Hirotaka and Sato, Shintaro},
    title = {Pauli String Partitioning Algorithm with the Ising
Model for Simultaneous Measurements},
    journal = {J. Phys. Chem. A},
    volume = {127},
    number = {4},
    pages = {1068-1080},
    year = {2023},
    month = {01},
    issn = {1089-5639},
    doi = {10.1021/acs.jpca.2c06453},
    url = {https://doi.org/10.1021/acs.jpca.2c06453},
    eprint = {https://pubs.acs.org/jpcafh/article-pdf/127/4/1068/2913745/jp2c06453.pdf},
}

@article{Bandyopadhyay2002,
    author = {Bandyopadhyay, Somshubhro and Boykin, P. Oscar and  Roychowdhury, Vwani and Vatan, Farrokh}, 
    title = {A New Proof for the Existence of Mutually Unbiased Bases}, 
    journal = {Algorithmica}, 
    volume = {34}, 
    pages = {512-528},
    year = {2002}, 
    month = {11}, 
    doi = {10.1007/s00453-002-0980-7}, 
    url = {https://doi.org/10.1007/s00453-002-0980-7}, 
}

@article{Kim2024,
    author = {Kim, Kyungmin and Lim, Sumin and Shin, Kyujin and Lee, Gwonhak and Jung, Yousung and Kyoung, Woomin and Rhee, June-Koo Kevin and Rhee, Young Min},
    title = {Variational quantum eigensolver for closed-shell molecules with non-bosonic corrections},
    journal = {Phys. Chem. Chem. Phys.},
    volume = {26},
    number = {10},
    pages = {8390-8396},
    year = {2024},
    month = {03},
    issn = {1463-9076},
    doi = {10.1039/d3cp05570a},
    url = {https://doi.org/10.1039/d3cp05570a},
}

@article{Alfonso2025,
    author = {Alfonso, Dominic and Lee, Yueh-Lin and Paudel, Hari P. and Duan, Yuhua},
    title = {Chemical applications of variational quantum eigenvalue-based quantum algorithms: Perspective and survey},
    journal = {Appl. Phys. Rev,},
    volume = {12},
    number = {3},
    pages = {031304},
    year = {2025},
    month = {07},
    issn = {1931-9401},
    doi = {10.1063/5.0245874},
    url = {https://doi.org/10.1063/5.0245874},
}

@article{OMalley2016,
  title = {Scalable Quantum Simulation of Molecular Energies},
  author = {O'Malley, P. J. J. and Babbush, R. and Kivlichan, I. D. and Romero, J. and McClean, J. R. and Barends, R. and Kelly, J. and Roushan, P. and Tranter, A. and Ding, N. and Campbell, B. and Chen, Y. and Chen, Z. and Chiaro, B. and Dunsworth, A. and Fowler, A. G. and Jeffrey, E. and Lucero, E. and Megrant, A. and Mutus, J. Y. and Neeley, M. and Neill, C. and Quintana, C. and Sank, D. and Vainsencher, A. and Wenner, J. and White, T. C. and Coveney, P. V. and Love, P. J. and Neven, H. and Aspuru-Guzik, A. and Martinis, J. M.},
  journal = {Phys. Rev. X},
  volume = {6},
  issue = {3},
  pages = {031007},
  numpages = {13},
  year = {2016},
  month = {Jul},
  publisher = {American Physical Society},
  doi = {10.1103/PhysRevX.6.031007},
  url = {https://link.aps.org/doi/10.1103/PhysRevX.6.031007}
}

@article{Clary2023,
author = {Clary, Jacob M. and Jones, Eric B. and Vigil-Fowler, Derek and Chang, Christopher and Graf, Peter},
title = {Exploring the scaling limitations of the variational quantum eigensolver with the bond dissociation of hydride diatomic molecules},
journal = {Int. J. Quantum Chem.},
volume = {123},
number = {11},
pages = {e27097},
doi = {https://doi.org/10.1002/qua.27097},
url = {https://onlinelibrary.wiley.com/doi/abs/10.1002/qua.27097},
year = {2023}
}

@article{Sugisaki2022,
    author = {Sugisaki, Kenji and Kato, Takumi and Minato, Yuichiro and Okuwaki, Koji and Mochizuki, Yuji},
    title = {Variational quantum eigensolver simulations with the multireference unitary coupled cluster ansatz: a case study of the C2v quasi-reaction pathway of beryllium insertion into a H2 molecule},
    journal = {Phys. Chem. Chem. Phys.},
    volume = {24},
    number = {14},
    pages = {8439-8452},
    year = {2022},
    month = {04},
    issn = {1463-9076},
    doi = {10.1039/d1cp04318h},
    url = {https://doi.org/10.1039/d1cp04318h}
}

@article{Tilly2020,
  title = {Computation of molecular excited states on IBM quantum computers using a discriminative variational quantum eigensolver},
  author = {Tilly, Jules and Jones, Glenn and Chen, Hongxiang and Wossnig, Leonard and Grant, Edward},
  journal = {Phys. Rev. A},
  volume = {102},
  issue = {6},
  pages = {062425},
  numpages = {9},
  year = {2020},
  month = {Dec},
  publisher = {American Physical Society},
  doi = {10.1103/PhysRevA.102.062425},
  url = {https://link.aps.org/doi/10.1103/PhysRevA.102.062425}
}

@article{Weaving2025,
  author  = {Weaving, Tim and Ralli, Alexis and Love, Peter J. and Succi, Sauro and Coveney, Peter V.},
  title   = {Contextual subspace variational quantum eigensolver calculation of the dissociation curve of molecular nitrogen on a superconducting quantum computer},
  journal = {npj Quantum Inf.},
  volume  = {11},
  pages   = {25},
  year    = {2025},
  doi     = {10.1038/s41534-024-00952-4},
  url     = {https://doi.org/10.1038/s41534-024-00952-4}
}

@article{Dalton2024,
  author  = {Dalton, Kieran and Long, Christopher K. and Yordanov, Yordan S. and Smith, Charles G. and Barnes, Crispin H. W. and Mertig, Normann and Arvidsson-Shukur, David R. M.},
  title   = {Quantifying the effect of gate errors on variational quantum eigensolvers for quantum chemistry},
  journal = {npj Quantum Inf.},
  volume  = {10},
  pages   = {18},
  year    = {2024},
  doi     = {10.1038/s41534-024-00808-x},
  url     = {https://doi.org/10.1038/s41534-024-00808-x}
}

\end{document}